\documentclass[12pt]{article}
\usepackage{amssymb,amsmath}

\begin{document}

\title{{\bf The Orbifolds of Permutation-Type as\\
Physical String Systems\\
 at Multiples of $\mathbf{c = 26}$\\
I.  Extended Actions and New Twisted World-Sheet Gravities}}
\author{M.~B. Halpern\footnote{halpern@physics.berkeley.edu}\\
Department of Physics, 
University of California, \\
and Theoretical Physics Group, \\
Lawrence Berkeley National Laboratory\\
University of California, 
Berkeley, California 94720, USA}

\maketitle

\begin{abstract}
\noindent This is the first in a series of papers in which I investigate the orbifolds
of permutation-type as candidates for {\it new physical string systems} at multiples of critical
closed-string central charges.  Examples include the bosonic orientation orbifolds,
the bosonic permutation orbifolds and others, as well as superstring extensions.
In this paper I use the extended (twisted) Virasoro algebras of these orbifolds
to construct the corresponding extended action formulations of the twisted
open- and closed-string sectors of all the bosonic orbifolds at ${\hat c} = 26K$.
The extended actions exhibit a large set of {\em new twisted world-sheet gravities},
whose extended diffeomorphism groups clearly indicate that the associated
operator-string theories can be free of negative-norm states at higher central
charge.
\end{abstract}

\clearpage
\tableofcontents

\clearpage
\section{Introduction}
\label{sec1}

At the level of examples, current-algebraic conformal field theory [1-11] and 
orbifold theory [11-22] are almost as old as string theory itself [23-26]. It is only
in the last few years however that the orbifold program [27-37] has in large 
part completed the local description of  the general closed-string 
orbifold conformal field theory $A(H)/H$. The program constructs all 
orbifolds at once, using the principle of local isomorphisms [29-32, 37] to map 
the $H$-symmetric CFT $A(H)$ into the twisted sectors of $A(H)/H$. For 
the reader interested in particular topics, the following list may be helpful:
\begin{itemize}
\item the twisted current algebras and stress tensors of the general 
current-algebraic orbifold [29-31],
\item the twisted affine-primary fields, twisted operator algebras and 
twisted KZ equations \footnote{An abelian twisted KZ equation for the 
inversion orbifold $x\rightarrow-x$ was given earlier in Ref. [38].} of all WZW  orbifolds [32,33,35,36],
\item the world-sheet action formulation of all WZW and coset orbifolds in terms of 
so-called group orbifold elements with diagonal monodromy [32-36],
\item the action formulation and twisted Einstein equations of a large 
class of sigma-model orbifolds [37],
\item the general free-bosonic avatars of these constructions on abelian $g$ and the 
explicit form of their twisted vertex operators [33,35,37].
\end{itemize}
A pedagogical review of the program is included in Ref. [36]. Recent 
progress at the level of characters has been reported in Refs. [27,39]. 
Complimentary discussions of WZW twist fields and twisted free-bosonic
vertex operators are found in Refs. [40,41] and [42] respectively.

In a second chapter, the technology of the orbifold program has also been 
applied to general twisted {\it open-string} conformal field theory:
\begin{itemize}
\item the WZW orientation orbifolds, including their branes and twisted 
open-string KZ equations [43],
\item the action formulation, branes and twisted Einstein equations of 
WZW, coset and sigma-model orientation orbifolds [44],
\item the open-string WZW orbifolds [45],
\item the general twisted open WZW string, including all 
$T$-dualizations, branes, non-commutative geometry and twisted 
open-string KZ systems [46],
\item the general twisted boundary-state equation of twisted open WZW 
strings [33,45,46],
\item free-bosonic avatars of these constructions on abelian $g$ [43,45,46].
\end{itemize}
The general orientation-orbifold CFT, which is constructed by twisting world-sheet 
orientation-reversing automorphisms, is apparently \footnote {I will have more to 
say about the relation between orientation orbifolds and orientifolds in 
succeeding papers of this series.} a twisted 
generalization of orientifold CFT [47]. The construction of the general 
twisted open WZW string is a synthesis of closed-string orbifold theory 
and the theory of untwisted open WZW strings given in Ref. [48]. Complementary
discussions of the untwisted open WZW string are found in Refs. [49,50], 
and there is 
also a complementary literature on twisted branes (see e.g. Ref. [51]) and
permutation branes [52] on group manifolds.

The present series of papers opens a third, more phenomenological chapter 
in the orbifold program -- in which some of the simplest conformal-field-theoretic 
results of the program are applied to physical string theory: My goal 
here is to ask whether simple {\it orbifolds of permutation-type} can describe 
 {\it new physical string systems} at higher central charge, for example the values
\renewcommand{\theequation}{\thesection.\arabic{equation}}
\setcounter{equation}{0}
\begin{equation}
\label{eq1.1}
{\hat c} = 26K,\quad K = 2,3,4,\dots
\end{equation}
which are found in the critical free-bosonic orbifolds of permutation-type.
These orbifolds begin with one or more copies of the critical closed bosonic 
string, so that (choosing the  closed-string critical dimension $d=26$) 
many examples of this type have already been studied at the level of 
orbifold conformal field theory:
\begin{itemize}
\item the twisted open-string sectors of the free-bosonic orientation 
orbifolds [43,44,46] at $\hat c=52$,
\item the twisted closed-string sectors of the free-bosonic permutation 
orbifolds [27,33,35,37] at any multiple of $c=26$,
\item the free-bosonic open-string permutation orbifolds  and their 
$T$-duals [45,46] at any multiple of $c=26$
\item the generalized [46] closed- and open-string permutation orbifolds at 
$\hat c=26K$, including extra automorphisms which act uniformly on each copy of 
the original closed string (see also Subsec. 4.2).
\end{itemize} 
The corresponding critical superstring orbifolds of permutation-type can also be studied
with central charges
\begin{equation}
\label{eq1.2}
{\hat c} = 10K,\quad K = 2,3,\dots
\end{equation}
or $({\hat c},{\hat{\bar c}}) = (26K,10K)$ for heterotic type. 
Here the subject is not as well developed, but some discussion of 
permutation-orbifold superconformal
field theory is found in Refs. [22,27]. 

The orbifolds of permutation-type have not previously been considered as candidates
for physical string systems. One reason for this may be that the covariant 
formulation of the twisted sectors exhibit extra twisted time-like 
currents\footnote{Untwisted theories with two time-like dimensions have 
been considered in Ref. [53].}, and hence extra sets of {\it negative 
norm states} (ghosts) associated with the higher central charge. It is 
important therefore to state the basic hypothesis which underlies this investigation:
``Orbifoldization should not create negative-norm states'' where there 
were none in the original symmetric theory. There are of course no ghosts 
in the untwisted sectors of these orbifolds -- which are after all nothing but 
(symmetrizations of) decoupled copies of ordinary ghost-free strings [54] 
-- and the orbifold program constructs the twisted sectors directly from the untwisted. On the
basis of this hypothesis then, we are led to expect a natural, extended mechanism
for ghost-decoupling in the twisted sectors of these orbifolds.

In this first installment of the series, I will find and study the 
classical precursor of this ghost-decoupling mechanism. The central 
observation is that each of the twisted sectors of these orbifolds contains
an {\it extended (twisted) Virasoro algebra} [27,55,35,43], which straightforwardly implies
a {\it new extended (twisted) world-sheet gravity} in each sector. 
Because the new extended gravities are in 1-1 correspondence with the conjugacy
classes of all the permutation groups, I will refer to them collectively 
as the {\it permutation gravities} (or $P$-gravities). The extended 
diffeomorphism groups of the permutation gravities clearly indicate that 
the corresponding twisted operator-string theories will exhibit new 
twisted BRST systems and new extended Ward identities -- so that their 
string amplitudes can be free of negative-norm states. These and other 
topics at the operator level will
be addressed in the succeeding papers of the series.

The world-sheet permutation gravities follow closely in the tradition of earlier
extended world-sheet gravities, all of which are associated to extended 
Virasoro algebras and new critical central charges. I mention in particular
a) the world-sheet supergravities [56] associated to superconformal 
algebras [57,58], b) the $W$-gravities [59] associated to $W$-algebras [60], and c) the exotic
world-sheet gravities of the generic affine-Virasoro constructions [61,62] associated
to $K$-conjugate (commuting) pairs of Virasoro algebras [2,10,11]. Similarly, one 
may expect  new world-sheet {\it permutation supergravities} associated to 
the extended, twisted superconformal algebras [27] of superstring orbifolds 
of permutation-type.

The organization of this paper is as follows. I will consider two classes 
of examples, the twisted open-string sectors of the orientation orbifolds 
in Sec. 2 and the twisted closed-string sectors of the permutation 
orbifolds in Sec. 3. For both classes, I use the extended Virasoro 
algebras to construct the classical extended Hamiltonian systems -- which then
straightforwardly imply the {\it extended action formulations} of Polyakov-type. The final forms 
of these actions, including identification of the {\it extended (twisted) 
metrics}, are found in Eqs. (2.52) and (3.28). The corresponding extended actions 
of Nambu-type are found in Subsecs. 2.8 and 3.4.

The open-string orientation-orbifold sectors are all governed by the simple
case of $\mathbb{Z}_2$-permutation gravity, but the derivation in this 
case is complicated by the need to follow the boundary conditions 
(branes) of the twisted open strings. The replacement of boundary 
conditions by monodromies simplifies the derivation for the closed-string sectors
of the permutation orbifolds, but here one encounters the systematics of the general world-sheet
permutation gravity. In this case, I have also been able to find a 
simpler, complementary derivation of the extended actions (see Subsec. 3.5) 
directly from the principle of local isomorphisms. For this reason, and 
because the permutation orbifolds are more familiar, the reader may wish 
to begin with Sec. 3.

For generality, the results are worked out first for the orbifold CFT's 
at $\hat c=Kd$, where $d$ is the number of free bosons in each copy of the untwisted 
closed-string CFT. The results for the critical orbifold-string theories 
at $d=26$ and $\hat c=26K$ are however easily obtained at any stage of 
the development and, using our quantitative knowledge of the permutation gravities,
I finally return in Sec. 4 to the conjectured properties of the critical 
orbifold-string theories at the operator level.

\section{${\mathbb Z}_2$-Permutation Gravity  \\
\hspace*{.3in} in the Orientation Orbifolds}
\label{sec2}

\subsection{Extended Virasoro Algebra\\
 \hspace*{.3in} and Extended Hamiltonian}
\label{sec2.1}

The general orientation orbifold [43,44,46] is constructed as
\setcounter{equation}{0}
\begin{equation}
\label{eq2.1}
\frac {A(H_-)}{H_-},\quad H_- = \,{\mathbb Z}_2\mbox{(world sheet)}\times H
\end{equation}
where $A(H_-)$ is any closed-string CFT with central charge $c$ and $H_-$ 
is any symmetry group which includes world-sheet orientation-reversing automorphisms.
Like orientifolds [47], each orientation orbifold contains an equal number 
of closed-and open-string sectors but, in contrast to orientifolds, the 
generic orientation-orbifold sector contains fractional modeing and the open-string sectors 
live at $\hat c=2c$. The closed-string sectors (associated to the 
orientation-preserving subgroup of $H_-$) form an ordinary (space-time) 
orbifold [27-37] by themselves.

I concentrate here on the twisted open-string sectors (associated to the orientation-reversing
automorphisms), each of which exhibits an order-two orbifold Virasoro 
algebra [27,55,35,43]
\renewcommand{\theequation}{\thesection.\arabic{subsection}\alph{equation}}
\setcounter{subsection}{2}
\setcounter{equation}{0}
\begin{eqnarray}
[{\hat L}_u( m + \tfrac {u}{2}),\ {\hat L}_v( n + \tfrac {v}{2})]  &= &( m-n+ \tfrac {u-v}{2}){\hat L}_{u+v}( m+n+ \tfrac {u+v}{2}) \label{eq2.2a} \\
&+ &\tfrac {\hat c}{12} ( m + \tfrac {u}{2}) (( m + \tfrac {u}{2})^2 - 1) \delta_{m+n+\tfrac {u+v}{2},0} \nonumber 
\end{eqnarray}
\begin{equation}
{\hat c} = 2c,\quad {\bar u} = 0,1. \label{eq2.2b}
\end{equation}
Here and below I assume the standard periodicity for all the spectral 
indices, so that $\bar u$ is the pullback of $u$ to the fundamental range 
-- and up and down indices $u$ are equivalent. The (integral) Virasoro subalgebra 
of the extended algebra (2.2) is generated by $\{L_0(m)\}$, and Sec. 2 of 
Ref. [43] gives a simple explanation of the transition $c\rightarrow \hat c=2c$ for 
the open-string sectors as well as the presence of the twisted generators
 $\left\{{\hat L}_1\left(m + \frac {1}{2}\right)\right\}$. In what 
 follows, I will use essentially standard methods [63,61] and the classical analogue of the extended Virasoro generators
  to construct the correspondingly-extended classical Hamiltonian of the general open-string 
orientation-orbifold sector. 
 
 I begin this discussion with the extended left- and right-mover stress 
 tensors of the sector
\setcounter{subsection}{3}
\setcounter{equation}{0}
\begin{equation}
{\hat \theta}_u^{\pm}(\xi,t) \equiv \tfrac {1}{2\pi} \sum_{m \in {\mathbb Z}}
 {\hat L}_u(m + \tfrac {u}{2}) e^{-i( m+\tfrac {u}{2})(t \pm \xi)},\quad {\bar u} = 0,1\label{eq2.3a} 
\end{equation}
\begin{equation}
{\hat \theta}_u^{\pm} (\xi + 2\pi,t) = (-1)^u{\hat \theta}_u^{\pm}(\xi,t) \label{eq2.3b} 
\end{equation}
\begin{equation}
{\hat \theta}_u^{\mp}(\xi,t) = {\hat \theta}_u^{\pm}(-\xi,t) \label{eq2.3c}
\end{equation}
both of which are constructed from the same, single set of extended Virasoro 
generators. These are the extended stress tensors of the 
 conformal field theory, whose form suffices to compute the following equal-time 
bracket algebra in the classical theory 
\setcounter{subsection}{4}
\setcounter{equation}{0}
\begin{eqnarray}
\{{\hat \theta}_u^+(\xi,t),{\hat \theta}_v^+(\eta,t)\} &= &i(\partial_{\xi} - \partial_{\eta})({\hat \theta}_{u+v}^+(\eta,t) \delta_{\frac {u}{2}}(\xi-\eta)) \label{eq2.4a} \\
&= &i\theta_{u+v}^+(\eta,t)\partial_{\xi}\delta_{\frac {u}{2}}(\xi-\eta)  \label{eq2.4b} \\
& &- i{\hat \theta}_{u+v}^+(\xi,t)\partial_{\eta}\delta_{\frac {v}{2}}(\eta-\xi) \nonumber \\
\delta_{\frac {u}{2}}(\xi-\eta) &= \!&\!\tfrac {1}{2\pi} \sum_{m \in {\mathbb Z}} e^{-i( m + \tfrac {u}{2})(\xi-\eta)} = e^{-i\frac {u}{2}(\xi-\eta)} \delta(\xi-\eta). \label{eq2.4c}
\end{eqnarray}
from the classical analogue of the extended Virasoro algebra. The quantity in Eq. (2.4c) is called a
 phase-modified delta function [37,45,46], 
and the other equal-time brackets of ${\{\hat \theta}_u^{\pm}\}$ follow from 
this result and Eq. (2.3c).

In terms of the classical extended stress tensors, one may define the 
classical {\it extended Hamiltonian} $\hat H$ of the sector, as well as the generator $\hat G$ of 
time-independent gauge transformations
\setcounter{subsection}{5}
\setcounter{equation}{0}
\begin{equation}
\label{eq2.5a}
{\hat H} \equiv \int_0^{2\pi} d\xi \sum_u {\hat v}_+^u(\xi,t) {\hat \theta}_u^+(\xi,t),\quad\, {\hat G} \equiv \int_0^{2\pi} d\xi \sum_u {\hat \epsilon}_+^u(\xi){\hat \theta}_u^+(\xi,t)
\end{equation}
\begin{equation}
\label{eq2.5b}
{\hat v}_+^u(\xi + 2\pi,t) = (-1)^u{\hat v}_+^u(\xi,t),\,\,\, {\hat \epsilon}_+^u(\xi + 2\pi) = (-1)^u{\hat \epsilon}_+^u(\xi),\,\,\,\, {\bar u} = 0,1
\end{equation}
where $\hat v$ and $\hat \epsilon$ are respectively the multipliers and 
the gauge parameters. Note that the densities here have trivial 
monodromy. We are interested however in the equivalent {\it open-string} forms 
of $\hat H$ and $\hat G$, which are integrated over the strip $0\leq\xi 
\leq\pi$:
\setcounter{subsection}{6}
\setcounter{equation}{0}
\begin{equation}
{\hat H} = \int_0^{\pi} d\xi \sum_u ({\hat v}_+^u(\xi,t){\hat \theta}_u^+(\xi,t) + {\hat v}_-^u(\xi,t){\hat \theta}_u^-(\xi,t)) \label{eq2.6a}
\end{equation}
\begin{equation}
{\hat G} = \int_0^{\pi} d\xi \sum_u ({\hat \epsilon}_+^u(\xi){\hat \theta}_u^+(\xi,t) + {\hat \epsilon}_-^u(\xi){\hat \theta}_u^-(\xi,t)) \label{eq2.6b}
\end{equation}
\begin{equation}
{\hat v}_-^u(\xi,t) = {\hat v}_+^u(-\xi,t),\quad {\hat \epsilon}_-^u(\xi) = {\hat \epsilon}_+^u(-\xi). \label{eq2.6c}
\end{equation}
In what follows, I will work directly with these open-string forms -- including the following
strip boundary conditions for all $n \in {\mathbb Z}_{\ge 0}$
\renewcommand{\theequation}{\thesection.\arabic{equation}}
\setcounter{equation}{6}
\begin{equation}
\label{eq2.7}
\partial_{\xi}^n {\hat v}_-^u(0,t) = (-1)^n\partial_{\xi}^n{\hat v}_+^u(0,t),\quad \partial_{\xi}^n{\hat v}_-^u(\pi,t) = (-1)^{n+u}\partial_{\xi}^n{\hat v}_+^u(\pi,t)
\end{equation}
which follow from Eqs. (2.5b) and (2.6c). The same boundary conditions 
hold for the parameters $\{{\hat \epsilon}_{\pm}^u\}$, and indeed for the 
stress tensors $\{{\hat \theta}_u^{\pm}\}$ themselves.

I turn now to the Hamiltonian equations of motion, beginning with the 
extended classical constraints
\begin{equation}
\label{eq2.8}
{\hat \theta}_u^{\pm}(\xi,t) = 0\,\,\, \leftrightarrow\,\,\, \{{\hat L}_u( m + \tfrac {u}{2}) = 0\},\quad {\bar u} = 0,1
\end{equation}
which are obtained by varying the multipliers in $\hat H$. These 
conditions generalize the standard classical Polyakov constraints ${\hat \theta}_0 =  {\hat L}_0(m) = 0$,
which are included here when $\bar u=0$. With these extended constraints 
included in the Hamiltonian formulation, we will not be surprised to 
recover them as well in the equivalent, extended action formulation (see 
Subsec. 2.7) of each twisted sector.

The other Hamiltonian equations of motion are specified as usual by
\begin{equation}
\label{eq2.9}
{\dot {\hat A}} = i\{H,A\}
\end{equation}
for all observables beyond the multipliers.  With the brackets (2.4), this 
defines the time-dependence of the {\it gauge-variant} stress tensors
\begin{equation}
\label{eq2.10}
{\dot {\hat \theta}}_u^{\pm} = \pm \sum_v [\partial_{\xi}({\hat \theta}_{u+v}^{\pm}{\hat v}_{\pm}^v) + {\hat \theta}_{u+v}^{\pm} \partial_{\xi} {\hat v}_{\pm}^v]
\end{equation}
which must reduce to the CFT stress tensors (2.3) in a certain gauge to 
be discussed below. Then the time dependence of the multipliers
\begin{equation}
\label{eq2.11}
{\dot {\hat v}}_{\pm}^u = \mp \sum_v {\hat v}_{\pm}^{u-v} {\stackrel{\leftrightarrow}{\partial}}_{\xi} {\hat v}_{\pm}^v,\quad
{\hat A} {\stackrel{\leftrightarrow}{\partial}}_{\xi}{\hat B} \equiv {\hat A} \partial_{\xi}{\hat B} - (\partial_{\xi}{\hat A}){\hat B}
\end{equation}
is obtained from Eq. (2.10) and the requirement that ${\dot {\hat H}} = 0$.

Similarly, the time-independent gauge transformations are specified as 
\renewcommand{\theequation}{\thesection.\arabic{subsection}\alph{equation}}
\setcounter{subsection}{12}
\setcounter{equation}{0}
\begin{eqnarray}
\delta {\hat A} &= &i\{{\hat G},{\hat A}\},\quad\, \delta {\hat v}_{\pm}^u \equiv \mp \sum_v{\hat v}_{\pm}^{u-v} {\stackrel{\leftrightarrow}{\partial}}_{\xi} {\hat \epsilon}_{\pm}^v \label{eq2.12a} \\
\delta {\hat \theta}_u^{\pm} &= &\pm \sum_v [\partial_{\xi}({\hat \theta}_{u+v}^{\pm} {\hat \epsilon}_{\pm}^v) + {\hat \theta}_{u+v}^{\pm} \partial_{\xi} {\hat \epsilon}_{\pm}^v]. \label{eq2.12b}
\end{eqnarray}
It follows that the extended Hamiltonian (2.6a) is {\it gauge-invariant}
\renewcommand{\theequation}{\thesection.\arabic{equation}}
\setcounter{equation}{12}
\begin{equation}
\label{eq2.13}
\delta {\hat H} = 0
\end{equation}
under the gauge group associated to the extended Virasoro algebra.

Among possible gauge conditions, I mention first the (partially-fixed) 
{\it Polyakov gauge}
\begin{equation}
\label{eq2.14}
{\hat v}_{\pm}^u = {\hat v}_{\pm}^0 \delta_{u,0 \!\!\!\!\mod 2} :\quad {\hat H} = \int_0^{\pi} d\xi({\hat v}_+^0{\hat \theta}_0^+ + {\hat v}_-^0{\hat \theta}_0^-)
\end{equation}
in which the Polyakov form of the Hamiltonian is recovered. On the other 
hand, the {\it conformal gauge}
\begin{equation}
\label{eq2.15}
{\hat v}_{\pm}^u = \delta_{u,0 \!\!\!\!\mod 2} :\quad {\hat H} = \int_0^{\pi} d\xi(\theta_0^+ + \theta_0^-) = {\hat L}_0(0)
\end{equation}
is a completely-fixed gauge which reproduces the extended stress tensors 
(2.3) and Hamiltonian of the twisted open-string CFT. In particular, the 
expected time dependence of the extended stress tensors
\begin{equation}
\label{eq2.16}
\partial_{\mp}{\hat \theta}_u^{\pm} = 0,\quad {\bar u} = 0,1,\,\,\, \partial_{\pm} \equiv \partial_t \pm \partial_{\xi}
\end{equation}
is obtained from Eq. (2.10) in the conformal gauge.

The corresponding extended action formulation of each open-string 
orientation orbifold sector can now be obtained by the standard Legendre transformation
\begin{equation}
\label{eq2.17}
{\hat S} = \int dt \int_0^{\pi} d\xi \sum_u ({\dot {\hat x}}^{n(r)\mu u} {\hat p}_{n(r)\mu u} - {\hat v}_+^u {\hat \theta}_u^+ - {\hat v}_-^u{\hat \theta}_u^-)
\end{equation}
from the phase-space sigma-model form of the extended stress tensors, 
where $\hat x$ and $\hat p$ are the twisted coordinates and momenta of 
each sector. Following the discussion of Refs. [63,61], we know that each 
extended action will exhibit a further-extended gauge invariance with 
world-sheet space- and {\it time}-dependent gauge parameters:
\renewcommand{\theequation}{\thesection.\arabic{subsection}\alph{equation}}
\setcounter{subsection}{18}
\setcounter{equation}{0}
\begin{equation}
\label{eq2.18a}
{\hat \epsilon}_{\pm}^u(\xi) \to {\hat \epsilon}_{\pm}^u(\xi,t),\quad {\bar u} = 0,1
\end{equation}
\begin{equation}
\label{eq2.18b}
\delta {\hat v}_{\pm}^u = 
{\dot {\hat \epsilon}}_{\pm}^u \mp \sum_v {\hat v}_{\pm}^{u-v} 
{\stackrel{\leftrightarrow}{\partial}}_{\xi} {\hat \epsilon}_{\pm}^v,\quad \delta {\hat x}^{n(r)\mu u} = \{{\hat G},{\hat x}^{n(r)\mu u}\}.
\end{equation}
These transformations are in fact a form of the  infinitesimal extended
diffeomorphisms of world-sheet {\it $\mathbb{Z}_2$-permutation gravity}, 
whose covariant form will be more transparent in coordinate space. To obtain an explicit form 
of the extended action however, we must choose a specific class of models.

\setcounter{subsection}{1}
\subsection{The Free-Bosonic Orientation Orbifolds}
\label{sec2.2}

Ref. [46] gives the phase-space sigma model description of the 
open-string sectors of the general WZW orientation orbifold, and it would 
be interesting to work out the extended actions \footnote{The 
CFT (or conformal gauge) action of each open-string  WZW orientation-orbifold sector
is known [44] in terms of group orbifold elements on the solid half 
cylinder} in this case. I confine the discussion here however to the 
simple case of the free-bosonic orientation orbifolds
\renewcommand{\theequation}{\thesection.\arabic{equation}}
\setcounter{equation}{18}
\begin{equation}
\label{eq2.19}
\frac {U(1)^d}{{\mathbb Z}_2(w.s.)\times H}
\end{equation}
which are also discussed in that reference as a special case on abelian 
$g$. In these cases, the twisted open-string sectors live at central charge 
$\hat c = 2d$.

In the notation of Eq. (2.19), the non-trivial element of ${\mathbb Z}_2(w.s.)$ 
permutes the left- and right-mover currents $J,\bar{J}$ of the untwisted
 closed-string CFT $U(1)^d$ 
while the extra automorphisms $H$ act uniformly on the left-and 
right-movers. More precisely, each open-string sector of the orientation orbifold
is obtained by twisting the action of a world-sheet orientation-reversing
automorphism
\renewcommand{\theequation}{\thesection.\arabic{subsection}\alph{equation}}
\setcounter{subsection}{20}
\setcounter{equation}{0}
\begin{equation}
[J_a(m),J_b(n)] = [{\bar J}_a(m), {\bar J}_b(m)] = mG_{ab}\delta_{m+n,0} \label{eq2.20a}
\end{equation}
\begin{equation}
J_a(m)' = \omega_a{}^b{\bar J}_b(m),\quad {\bar J}_a(m)' = \omega_a{}^b J_b(m) \label{eq2.20b} 
\end{equation}
\begin{equation}
\omega_a{}^c\omega_b{}^dG_{cd} = G_{ab},\,\,\, \omega \in H \label{eq2.20c}
\end{equation}
\begin{equation} 
m,n \in {\mathbb Z},\quad a,b = 1\dots d. \label{eq2.20d}
\end{equation}
where the quantity $G_{ab}$ is the tangent-space metric of the untwisted 
CFT, with inverse $G^{ab}$. At any stage in the discussion below, one may 
substitute the explicit form
\renewcommand{\theequation}{\thesection.\arabic{equation}}
\setcounter{equation}{20}
\begin{equation}
\label{eq2.21}
\frac {U(1)^{26}}{{\mathbb Z}_2(w.s.) \times H} :\quad G_{ab} = G^{ab} = 
\begin{pmatrix} -1 & 0 \\ 0 & 1\!\!1 \end{pmatrix},\,\, a,b = 0,1,\dots 25
\end{equation}
to obtain the results for the orientation orbifolds of the critical 
Minkowski-space closed string. In this case of course, all the twisted 
open-string sectors live at operator central charge $\hat c = 52$.

For each of these twisted open-string sectors, Ref. [46] gives 
the form of the classical extended stress tensors in terms of the twisted 
currents of each sector:
\renewcommand{\theequation}{\thesection.\arabic{subsection}\alph{equation}}
\setcounter{subsection}{22}
\setcounter{equation}{0}
\begin{equation}
{\hat \theta}_u^{\pm}(\xi) = \tfrac {1}{8\pi} {\mathcal G}^{n(r)\mu;n(s)\nu} \sum_v {\hat J}_{n(r)\mu v}^{\pm}(\xi) {\hat J}_{n(s)\nu,u-v}^{\pm}(\xi),\quad {\bar u} = 0,1 \label{eq2.22a}
\end{equation}
\begin{eqnarray}
\{{\hat J}_{n(r)\mu u}^+(\xi),{\hat J}_{n(s)\nu v}^+(\eta)\} &= &4\pi i \delta_{n(r)+n(s),0 \!\!\!\!\mod\rho(\sigma)} \delta_{u+v,0 \!\!\!\!\mod 2} \label{eq2.22b} \\
& &\times {\mathcal G}_{n(r)\mu;-n(r),\nu} \partial_{\xi} \delta_{\frac {n(r)}{\rho(\sigma)} + \tfrac {u}{2}} (\xi-\eta) \nonumber
\end{eqnarray}
\begin{equation}
{\hat J}_{n(r)\mu u}^{\pm}(\xi,t) = \sum_m {\hat J}_{n(r)\mu u} (m + \tfrac {n(r)}{\rho(\sigma)} + \tfrac {u}{2}) e^{-i(m + \tfrac {n(r)}{\rho(\sigma)} + \tfrac {u}{2})(t\pm \xi)} \label{eq2.22c} 
\end{equation}
\begin{equation}
{\hat J}_{n(r)\mu u}^-(0,t) = {\hat J}_{n(r)\mu u}^+(0,t),\,\,\,\, {\hat J}_{n(r)\mu u}^-(\pi,t) = e^{2\pi i(\tfrac {n(r)}{\rho(\sigma)} + \tfrac {u}{2})} {\hat J}_{n(r)\mu u}^+(\pi,t). \label{eq2.22d} 
\end{equation}
Here the quantity
\renewcommand{\theequation}{\thesection.\arabic{equation}}
\setcounter{equation}{22}
\begin{eqnarray}
\label{eq2.23}
{\mathcal G}_{n(r)\mu;n(s)\nu} &= &\chi_{n(r)\mu}\chi_{n(s)\nu} U_{n(r)\mu}{}^a U_{n(s)\mu}{}^b G_{ab} \\
&= &\delta_{n(r)+n(s),0 \!\!\!\!\mod \rho(\sigma)} {\mathcal G}_{n(r)\mu;-n(r),\nu} \nonumber
\end{eqnarray}
is the twisted tangent-space metric of the sector, which is a  
duality transformation [29,31,32,37] of the untwisted metric. The quantities 
$\{\chi\}$ are normalization constants and $\mathcal 
G^.$ in the extended stress tensors is the inverse of $\mathcal G.$. The 
unitary eigenmatrices $U$ in the duality transformation (2.23) are 
determined by the so-called $H$-eigenvalue problem [29,31,32,37] of each automorphism $\omega$ 
(see Eq. (2.20)) 
\renewcommand{\theequation}{\thesection.\arabic{subsection}\alph{equation}}
\setcounter{subsection}{24}
\setcounter{equation}{0}
\begin{equation}
\omega_a{}^b(U^{\dag})_b{}^{n(r)\mu} = (U^{\dag})_a{}^{n(r)\mu} e^{-2\pi i \frac {n(r)}{\rho(\sigma)}},\quad \omega \in H \label{eq2.24a}
\end{equation}
\begin{equation} 
{\bar n}(r) \in (0,1,\dots \rho(\sigma)-1) \label{eq2.24b}
\end{equation}
where $\rho(\sigma), n(r),\mu$ are the order and the
spectral and degeneracy indices respectively of $\omega$. Following convention in the 
orbifold program, all quantities are periodic $n(r) \to n(r) \pm 
\rho(\sigma)$ in the spectral indices, $\bar n(r)$ is the pullback of 
$n(r)$ to the fundamental region, and I have suppressed explicit sums 
over repeated indices $\{\bar n(r),\mu\}$ in the stress tensors.

For the phase-space formulation, we also need the quasi-canonical 
realization of the twisted currents [46]
\renewcommand{\theequation}{\thesection.\arabic{subsection}\alph{equation}}
\setcounter{subsection}{25}
\setcounter{equation}{0}
\begin{equation}
{\hat J}_{n(r)\mu u}^+ = 2\pi {\hat p}_{n(r)\mu u} + {\mathcal G}_{n(r)\mu;n(s)\nu} \partial_{\xi} {\hat x}^{n(s)\nu,-u} \label{eq2.25a}
\end{equation}
\begin{equation}
{\hat J}_{n(r)\mu u}^- = (-1)^{u+1}(2\pi {\hat p}_{n(r)\mu u} - {\mathcal G}_{n(r)\mu;n(s)\nu} \partial_{\xi} {\hat x}^{n(s)\nu,-u}) \label{eq2.25b}
\end{equation}
where $\{{\hat x}^{n(r)\mu u}\}$ and $\{{\hat p}_{n(r)\mu u}\}$ are the 
extended, twisted coordinates and momenta. Because of the extra label $\bar u = 
0,1$, each twisted open-string sector has exactly $2d$ extended 
coordinates \footnote {As an example, the automorphism 
$\omega=-1$ gives $\rho(\sigma)=2, U=1, \bar{n}=1$ and $\mu=a$, so that this 
twisted sector has d 
coordinates $\{\hat x^{1a0}\}$ with $DN$ boundary conditions and d 
coordinates $\{\hat x^{1a1}\}$ with $NN$ boundary conditions. This and many 
other examples are further discussed in Refs. [43,44,46].} in agreement with the 
 central charge $\hat c =2d$ of each of these sectors. The 
complete quasi-canonical algebra of $\hat x$ and $\hat p$ (including the 
twisted non-commutative geometry of $\hat x$ with itself) is given in Eq. 
(4.33) of Ref. [46], but we will need here only the brackets of the twisted 
coordinates with the currents [46]:
\renewcommand{\theequation}{\thesection.\arabic{subsection}\alph{equation}}
\setcounter{subsection}{26}
\setcounter{equation}{0}
\begin{eqnarray}
& &\hspace{-.2in}\{{\hat J}_{n(r)\mu u}^+(\xi),{\hat x}^{n(s)\nu v}(\eta)\} \label{eq2.26a} \\
& &\qquad\hspace{-.4in}= -2\pi i \delta_{n(r)\mu u}{}^{n(s)\nu v} (\delta_{{\bar y}(r,u)}(\xi-\eta) + (-1)^{u+1} \delta_{{\bar y}(r,u)}(\xi+\eta))\nonumber
\end{eqnarray}
\begin{eqnarray}
& &\hspace{-.2in}\{{\hat J}_{n(r)\mu u}^-(\xi), {\hat x}^{n(s)\nu v}(\eta)\} \label{eq2.26b} \\
& &\qquad\hspace{-.4in} = -2\pi i \delta_{n(r)\mu u}{}^{n(s)\nu v} ((-1)^{u+1} \delta_{-{\bar y}(r,u)}(\xi-\eta) + \delta_{-{\bar y}(r,u)}(\xi+u))\nonumber 
\end{eqnarray}
\begin{equation}
{\bar y}(r,u) = \tfrac {{\bar n}(r)}{\rho(\sigma)} + \tfrac {\bar u}{2}. \label{eq2.26c}
\end{equation}
These brackets, the extended Hamiltonian (2.6a) and the Hamiltonian 
equations of motion (2.9) suffice to compute the derivatives of the 
twisted coordinates:
\renewcommand{\theequation}{\thesection.\arabic{subsection}\alph{equation}}
\setcounter{subsection}{27}
\setcounter{equation}{0}
\begin{equation}
{\dot {\hat x}}^{n(r)\mu u} = \tfrac {1}{2} {\mathcal G}^{n(r)\mu;n(s)\nu} \sum_v ({\hat v}_+^v {\hat J}_{n(s)\nu,v-u}^+ 
 + (-1)^{u+1} {\hat v}_-^v{\hat J}_{n(s)\nu,v-u}^-) \label{eq2.27a} 
 \end{equation}
 \begin{equation}
\partial_{\xi} {\hat x}^{n(r)\mu u} = \tfrac {1}{2} {\mathcal G}^{n(r)\mu;n(s)v} \sum_v ({\hat J}_{n(s)\nu,-u}^+ + (-1)^u {\hat J}_{n(s)\nu,-u}^-). \label{eq2.27b}
\end{equation}
The spatial derivative in (2.27b) follows easily from the phase-space 
realization of the twisted currents.

Finally, I give the extended infinitesimal gauge transformation of the 
twisted coordinates
\renewcommand{\theequation}{\thesection.\arabic{equation}}
\setcounter{equation}{27}
\begin{equation}
\delta{\hat x}^{n(r)\mu u} = \frac {1}{2} {\mathcal G}^{n(r)\mu;n(s)\nu} \sum_v ({\hat \epsilon}_+^v {\hat J}_{n(s)\nu,v-u}^+ + (-1)^{u+1} {\hat \epsilon}_-^v {\hat J}_{n(s)\nu,v-u}^-)
\end{equation}
which follows (in parallel to $\dot {\hat x}$) from Eq. (2.12a).

\setcounter{subsection}{2}
\subsection{Coordinate-Space, Branes and Twisted Currents}
\label{sec2.3}

We may now begin the passage to coordinate space, where world-sheet 
$\mathbb{Z}_2$-permutation gravity can be seen in covariant form.

I begin with the boundary conditions or {\it branes} at the ends of the twisted open strings
\renewcommand{\theequation}{\thesection.\arabic{subsection}\alph{equation}}
\setcounter{subsection}{29}
\setcounter{equation}{0}
\begin{equation}
\label{eq2.29a}
{\dot {\hat x}}^{n(r)\mu 0}(0) = \partial_{\xi} {\hat x}^{n(r)\mu 1}(0) = 0
\end{equation}
\begin{equation}
\label{eq2.29b}
\cos(\tfrac {n(r)\pi}{\rho(\sigma)}) {\dot {\hat x}}^{n(r)\mu u}(\pi) - i \sin (\tfrac {n(r)\pi}{\rho(\sigma)}) \sum_v {\hat v}_+^v(\pi) \partial_{\xi} {\hat x}^{n(r)\mu,u-v}(\pi) = 0
\end{equation}
which follow from Eq. (2.27) and the boundary conditions (2.7),(2.22d) on the 
multipliers and the currents. Not surprisingly, the branes are quite 
different at the two ends of the twisted string -- and I note in particular that the 
branes at $\pi$ depend on the boundary values of the multipliers (see 
also Eq. (2.37e)).

To move further into coordinate space, one needs to solve for the momenta 
in terms of the time derivatives. The required result
\renewcommand{\theequation}{\thesection.\arabic{subsection}\alph{equation}}
\setcounter{subsection}{30}
\setcounter{equation}{0}
\begin{equation}
{\hat p}_{n(r)\mu u} = \tfrac {1}{\pi} {\mathcal G}_{n(r)\mu;n(s)\nu} \sum_v (M^{-1}\partial_t - N \partial_{\xi})_{uv} {\hat x}^{n(s)\nu v} \label{eq2.30a}
\end{equation}
\begin{equation}
M^{uv} = M^{(u+v)},\ M^{(w)} \equiv {\hat v}_+^w + (-1)^w{\hat v}_-^w \label{eq2.30b}
\end{equation}
\begin{equation}
(M^{-1})_{uv} = (M^{-1})_{(u+v)},\ M_{(w)}^{-1} = \gamma^{-1} ({\hat v}_-^w + (-1)^w {\hat v}_+^w) \label{eq2.30c} 
\end{equation}
\begin{equation}
N_{uv} = N_{(u+v)} \label{eq2.30d} 
\end{equation}
\begin{eqnarray}
2N_{(0)} &\equiv &\gamma^{-1} (({\hat v}_+^0)^2 + ({\hat v}_-^1)^2 - ({\hat v}_-^0)^2 - ({\hat v}_+^1)^2), \label{eq2.30e} \\
2N_{(1)} &\equiv &{\hat v}_+^1{\hat v}_-^0 + {\hat v}_-^1{\hat v}_+^0 \nonumber 
\end{eqnarray}
\begin{equation}
\gamma \equiv\,\, ({\hat v}_-^0 + {\hat v}_+^0)^2 - ({\hat v}_-^1 - {\hat v}_+^1)^2 \label{eq2.30f}
\end{equation}
follows after some algebra from Eqs. (2.25) and (2.27). The additional relations
\renewcommand{\theequation}{\thesection.\arabic{equation}}
\setcounter{equation}{30}
\begin{equation}
\label{eq2.31}
\sum_w M^{(w)}N_{(w+u)} = \frac {1}{2} ({\hat v}_+^u + (-1)^{u+1} {\hat 
v}_-^u),\quad {\bar u} =0,1
\end{equation}
are then obtained from the explicit form of the matrices $M$ and $N$.

As a first application of Eq. (2.30), the coordinate-space form of the 
twisted currents
\renewcommand{\theequation}{\thesection.\arabic{subsection}\alph{equation}}
\setcounter{subsection}{32}
\setcounter{equation}{0}
\begin{equation}
{\hat J}_{n(r)\mu u}^+ = 2{\mathcal G}_{n(r)\mu;n(s)\nu} \sum_v \left(M^{-1}\partial_t - (N - \tfrac {1}{2}) \partial_{\xi}\right)_{uv} {\hat x}^{n(s)\nu v} \label{eq2.32a}
\end{equation}
\begin{equation}
{\hat J}_{n(r)\mu u}^- = 2(-1)^{u+1} {\mathcal G}_{n(r)\mu;n(s)\nu} \sum_v \left( M^{-1} \partial_t - (N + \tfrac {1}{2}) \partial_{\xi}\right)_{uv}  {\hat x}^{n(r)\nu v}\label{eq2.32b}
\end{equation}
is easily obtained from the phase-space realization (2.25).

\setcounter{subsection}{3}
\subsection{Extended Polyakov Action\\
\hspace*{.5in}and ${\mathbb Z}_2$-twisted Permutation Gravity}
\label{sec2.4}

To obtain the explicit coordinate-space form of the extended Polyakov 
action (2.17) of these sectors, it is simplest to first express all 
quantities in terms of the twisted currents
\renewcommand{\theequation}{\thesection.\arabic{subsection}\alph{equation}}
\setcounter{subsection}{33}
\setcounter{equation}{0}
\begin{equation}
{\hat p}_{n(r)\mu u} = \tfrac {1}{4\pi} ({\hat J}_{n(r)\mu u}^+ + (-1)^{u+1} {\hat J}_{n(r)\mu u}^-) \label{eq2.33a}
\end{equation}
\begin{equation} 
{\hat S} = \int dt \int_0^{\pi} d\xi\, {\hat {\mathcal L}}_0 \label{eq2.33b} 
\end{equation}
\begin{equation}
\hspace{-1.5in}{\hat {\mathcal L}}_0 = \sum_u ({\dot {\hat x}}^{n(r)\mu u} {\hat p}_{n(r)\mu u} - {\hat v}_+^u {\hat \theta}_u^+ - {\hat v}_-^u {\hat \theta}_u^-) \label{eq2.33c} 
\end{equation}
\begin{equation}
\hspace{.7in}= \tfrac {1}{8\pi} {\mathcal G}^{n(r)\mu;n(s)\nu} \sum_{u,v} (-1)^{v+1} ({\hat v}_+^u{\hat J}_{n(r)\mu v}^- {\hat J}_{n(s)\nu,u-v}^+ 
 + (+\leftrightarrow -)) \label{eq2.33d}
\end{equation}
and then use the coordinate-space form (2.32) of the currents.

It is then straightforward to put the extended Polyakov action in the 
following ``covariant'' form
\renewcommand{\theequation}{\thesection.\arabic{subsection}\alph{equation}}
\setcounter{subsection}{34}
\setcounter{equation}{0}
\begin{equation}
{\hat S} = \tfrac {1}{4\pi} \int dt \int_0^{\pi} d\xi \sum_{u,v} {\tilde h}_{uv}^{mn} \partial_m {\hat x}^{n(r)\mu u} {\mathcal G}_{n(r)\mu;n(s)\nu} \partial_n{\hat x}^{n(s)\nu v} \label{eq2.34a} 
\end{equation}
\begin{equation}
\partial_m = (\partial_0,\partial_1) = (\tfrac {\partial}{\partial t}, \tfrac {\partial}{\partial \xi}),\quad\,\, {\tilde h}_{uv}^{mn} = {\hat h}_{(u+v)}^{mn} \label{eq2.34b}
\end{equation}
\begin{equation}
{\tilde h}_{(w)}^{00} \equiv 2M_{(w)}^{-1},\quad\,\, {\tilde h}_{(w)}^{10} = {\tilde h}_{(w)}^{01} \equiv -2N_{(w)} \label{eq2.34c} 
\end{equation}
\begin{equation}
{\tilde h}_{(w)}^{11} \equiv -\tfrac {1}{2} M^{(-w)} + \sum_u ({\hat v}_+^u + (-1)^{u+1} {\hat v}_-^u)N_{(w+u)} \label{eq2.34d}
\end{equation}
where the matrices $M$ and $N$ are defined in (2.30) and the identity (2.31) was 
used to simplify these results. All dependence on the four phase-space 
multipliers $\{{\hat v}_{\pm}^u,{\bar u} = 0,1\}$ is now collected in 
what I will call (and later motivate) the {\it extended inverse metric 
density} of $\mathbb{Z}_2$-permutation gravity
\renewcommand{\theequation}{\thesection.\arabic{equation}}
\setcounter{equation}{34}
\begin{equation}
\label{eq2.35}
{\tilde h}_{uv}^{mn} = {\tilde h}_{uv}^{nm} = \begin{pmatrix} {\tilde h}_{(0)}^{mn} & {\tilde h}_{(1)}^{mn} \\ {\tilde h}_{(1)}^{mn} & {\tilde h}_{(0)}^{mn} \end{pmatrix},\quad m,n \in (0,1)
\end{equation}
which then has {\it four} independent degrees of freedom.

As a check on the algebra, I record our coordinate-space results in the 
completely-fixed conformal gauge (2.15)
\renewcommand{\theequation}{\thesection.\arabic{subsection}\alph{equation}}
\setcounter{subsection}{36}
\setcounter{equation}{0}
\begin{equation}
{\hat v}_{\pm}^u = \delta_{u,0 \!\!\!\!\mod 2},\quad M_{(u)}^{-1} = \tfrac {1}{2} \delta_{u,0 \!\!\!\!\mod 2},\quad N_{(u)} = 0 \label{eq2.36a} 
\end{equation}
\begin{equation}
{\tilde h}_{(0)}^{mn} = \eta^{mn} = \begin{pmatrix} 1 & 0 \\ 0 & -1 \end{pmatrix},\quad {\tilde h}_{(1)}^{mn} = 0 \label{eq2.36b} 
\end{equation}
\begin{equation}
{\hat S} = \tfrac {1}{4\pi} \int dt \int_0^{\pi} d\xi \eta^{mn}{\mathcal G}_{n(r)\mu;n(s)\nu} \sum_u \partial_m {\hat x}^{n(r)\mu u} \partial_n{\hat x}^{n(s)\nu,-u} \label{eq2.36c} 
\end{equation}
\begin{equation}
{\dot {\hat x}}^{n(r)\mu 0}(0) = \partial_{\xi} {\hat x}^{n(r)\mu 1}(0) = 0 \label{eq2.36d} 
\end{equation}
\begin{equation}
{\dot {\hat x}}^{n(r)\mu u}(\pi) = i\tan (\tfrac {n(r)\pi}{\rho(\sigma)}) \partial_{\xi} {\hat x}^{n(r)\mu u}(\pi) \label{eq2.36e} 
\end{equation}
\begin{equation}
{\hat J}_{n(r)\mu u}^+ = {\mathcal G}_{n(r)\mu;n(s)\nu}\partial_+ 
{\hat x}^{n(r)\mu,-u} \label{eq2.36f}
\end{equation}
\begin{equation} 
 {\hat J}_{n(r)\mu u}^- = (-1)^{u+1} {\mathcal G}_{n(r)\mu;n(s)\nu} \partial_- {\hat x}^{n(r)\mu,-u} \label{eq2.36g}
\end{equation}
\begin{equation}
{\hat \theta}_u^+ = \tfrac {1}{8\pi} {\mathcal G}_{n(r)\mu;n(s)\nu} \sum_v \partial_+ {\hat x}^{n(r)\mu v} \partial_+ {\hat x}^{n(s)\mu,-u-v} \label{eq2.36h} 
\end{equation}
\begin{equation}
{\hat \theta}_u^- = \tfrac {1}{8\pi} (-1)^{u+1} {\mathcal G}_{n(r)\mu;n(s)\nu} \sum_v \partial_- {\hat x}^{n(r)\mu v} \partial_- {\hat x}^{n(r)\mu,-u-v} \label{eq2.36i} 
\end{equation}
\begin{equation}
\square {\hat x}^{n(r)\mu u} = 0,\quad \partial_{\mp} {\hat J}_{n(r)\mu u}^{\pm} = \partial_{\mp} {\hat \theta}_u^{\pm} = 0 \label{eq2.36j}
\end{equation}
where $\eta$ is the flat world-sheet metric. This is exactly the 
description given in Refs. [43,44,46] for the twisted open-string CFT's of the 
orientation orbifolds. Mixed boundary conditions such as Eq. (2.36e), even at 
vanishing twisted $B$ field, are well-known [43-46] in the orbifold program. The twisted 
mode expansions of the extended coordinates $\{\hat x\}$, as well as their 
twisted non-commutative geometries, are also given for the CFT's in Ref. [46].

As another check on the algebra, note that the extended action reduces in 
the Polyakov gauge (2.14) to the ordinary Polyakov action [64] of each twisted sector
\renewcommand{\theequation}{\thesection.\arabic{subsection}\alph{equation}}
\setcounter{subsection}{37}
\setcounter{equation}{0}
\begin{equation}
{\hat v}_{\pm}^1 = {\tilde h}_{(1)}^{mn} = 0 \label{eq2.37a} 
\end{equation}
\begin{equation}
{\tilde h}_{(0)}^{mn} = \sqrt{-h_{(0)}}\, h_{(0)}^{mn} = \tfrac {1}{{\hat v}_+^0 + {\hat v}_-^0} \begin{pmatrix} 2 & {\hat v}_-^0 - {\hat v}_+^0 \\ {\hat v}_-^0 - {\hat v}_+^0 & -2{\hat v}_+^0{\hat v}_-^0 \end{pmatrix} \label{eq2.37b}
\end{equation}
\begin{equation}
h_{mn}^{(0)} \equiv \mbox{sign}({\hat v}_-^0 + {\hat v}_+^0)e^{-\phi} \begin{pmatrix} {\hat v}_-^0{\hat v}_+^0 & \tfrac {1}{2}({\hat v}_-^0-{\hat v}_+^0) \\ \tfrac {1}{2}({\hat v}_-^0-{\hat v}_+^0) & -1 \end{pmatrix} \label{eq2.37c} 
\end{equation}
\begin{equation}
h_{mp}^{(0)}h_{(0)}^{pn} = \delta_m^n,\quad h_{(0)} \equiv \det(h_{mn}^{(0)}) \label{eq2.37d} 
\end{equation}
\begin{equation}
{\hat S} = \tfrac {1}{4\pi} \int dt \int_0^{\pi} d\xi \sqrt{-h_{(0)}}\, h_{(0)}^{mn} {\mathcal G}_{n(r)\mu;n(s)\nu} \sum_u \partial_m  {\hat x}^{n(r)\mu u}  \partial_n {\hat x}^{n()\nu,-u} \label{eq2.37e} 
\end{equation}
where $h_{mn}^{(0)}$ is the Polyakov metric and $\phi$ is the Weyl degree 
of freedom. It is then clear that the ``extended inverse metric density'' ${\tilde h}_{uv}^{mn}$ 
in Eq. (2.35) is a ``two-component'' extension of the inverse metric density $\sqrt{-h_{(0)}}\, h_{(0)}^{mn}$
of ordinary world-sheet gravity.

\setcounter{subsection}{4}
\subsection{Extended ($Z_2$-twisted) Diffeomorphisms}
\label{sec2.5}

I turn next to the coordinate-space form of the extended diffeomorphisms 
of $\mathbb{Z}_2$-permutation gravity.

After some algebra, one finds that the extended Hamiltonian gauge 
transformations (2.18b) and (2.28) can be put in the covariant form:
\renewcommand{\theequation}{\thesection.\arabic{subsection}\alph{equation}}
\setcounter{subsection}{38}
\setcounter{equation}{0}
\begin{equation}
\delta{\hat x}^{n(r)\mu u} = \sum_v {\hat \beta}^{mv} \partial_m {\hat x}^{n(r)\mu,u-v},\quad {\bar u} = 0,1 \label{eq2.38a}
\end{equation}
\begin{equation} 
\delta{\tilde h}_{(u)}^{mn} = \sum_v \{\partial_p ({\hat \beta}^{pv} {\tilde h}_{(u+v)}^{mn}) - (\partial_p{\hat \beta}^{mv}){\tilde h}_{(u+v)}^{pn} 
 - (\partial_p {\hat \beta}^{nv}) {\tilde h}_{(u+v)}^{pm}\}. \label{eq2.38b}
\end{equation}
The explicit form of the four extended diffeomorphism parameters $\{{\hat 
\beta}^{mu}, {\bar u} = 0,1\}$ is
\renewcommand{\theequation}{\thesection.\arabic{subsection}\alph{equation}}
\setcounter{subsection}{39}
\setcounter{equation}{0}
\begin{eqnarray}
{\hat \beta}^{tu} &\equiv &\sum_v ({\hat \epsilon}_+^v + (-1)^v {\hat \epsilon}_-^v)M_{(v-u)}^{-1},\quad {\bar u} = 0,1 \label{eq2.39a} \\
{\hat \beta}^{\xi u} &\equiv &\tfrac {1}{2} ({\hat \epsilon}_+^u + (-1)^{u+1} {\hat \epsilon}_-^u) - \sum_v ({\hat \epsilon}_+^v + (-1)^v {\hat \epsilon}_-^v) N_{(v-u)} \label{eq2.39b}
\end{eqnarray}
where the $\hat\epsilon$'s are the four gauge parameters of the extended 
Hamiltonian formulation. I remind that the $\hat\epsilon$'s and hence the 
$\hat\beta$'s have arbitrary world-sheet space-time dependence, thus 
comprising a true, {\it twisted doubling} of the standard world-sheet gravitational 
gauge degrees of freedom.

Using Eqs. (2.33b), (2.34a) and (2.38), one then obtains the 
corresponding transformation of the extended action
\renewcommand{\theequation}{\thesection.\arabic{subsection}\alph{equation}}
\setcounter{subsection}{40}
\setcounter{equation}{0}
\begin{eqnarray}
\delta {\hat {\mathcal L}}_0 &= &\partial_m \left( \sum_w {\hat \beta}^{mw}{\hat {\mathcal L}}_w\right) \label{eq2.40a} \\
{\hat {\mathcal L}}_w &\equiv &\tfrac {1}{4\pi} \sum_{u,v} {\tilde h}_{(w+u+v)}^{mn} \partial_m {\hat x}^{n(r)\mu u} {\mathcal G}_{n(r)\mu;n(s)\nu} \partial_n{\hat x}^{n(s)\nu v} \label{eq2.40b} \\
\delta {\hat S} &= &\int dt \sum_w ({\hat \beta}^{\xi w}(\pi){\hat 
{\mathcal L}}_w(\pi) - {\hat \beta}^{\xi w}(0){\hat {\mathcal L}}_w(0)) \label{eq2.40c}
\end{eqnarray}
where ${\hat {\mathcal L}}_w$ is the natural two-component extension of the 
action density  ${\hat {\mathcal L}}_0$. The result in Eq. (2.40c) 
requires that we study the coordinate-space boundary conditions in further detail.

The following boundary conditions on the $\mathbb{Z}_2$-gravitational structures
\renewcommand{\theequation}{\thesection.\arabic{subsection}\alph{equation}}
\setcounter{subsection}{41}
\setcounter{equation}{0}
\begin{eqnarray}
M_{(1)}^{-1}(0) &= &N_{(0)}(0) = N_{(u)}(\pi) = 0 \label{eq2.41a} \\
{\tilde h}_{(1)}^{00}(0) &= &{\tilde h}_{(1)}^{11}(0) = {\tilde h}_{(0)}^{01}(0) = 0 \label{eq2.41b} \\
{\hat \beta}^{\xi 0}(0) &= &{\hat \beta}^{\xi u}(\pi) = \beta^{t1}(0) = 0 \label{eq2.41c} \\
{\tilde h}_{(u)}^{11}(\pi) &= &-{\hat v}_+^u(\pi),\quad {\tilde h}_{(u)}^{01}(\pi) = 0 \label{eq2.41d}
\end{eqnarray}
follow from  the original Hamiltonian boundary 
conditions (2.7) on $\hat v$ and $\hat\epsilon$, using
the definitions of $M,N,\tilde h$ and $\beta$ in  Eqs. (2.30), 
(2.34) and (2.39). These are not quite enough for the invariance of $\hat 
S$, but these conditions and the boundary conditions (2.29a) on the twisted 
coordinates at $\xi = 0$ suffice to show that
\renewcommand{\theequation}{\thesection.\arabic{equation}}
\setcounter{equation}{41}
\begin{equation}
\label{eq2.42}
{\hat {\mathcal L}}_1(0) = 0\,\,\, \to\,\,\, \delta {\hat S} = 0.
\end{equation}
As expected from the Hamiltonian formulation, the extended action $\hat 
S$ is invariant under the extended ($\mathbb{Z}_2$-twisted) 
diffeomorphisms of $\mathbb{Z}_2$-permutation gravity.

\setcounter{subsection}{5}
\subsection{The Twisted Coordinate Equation of Motion}
\label{sec2.6}

It is instructive to vary the extended action (2.34a) by arbitrary 
infinitesimal variations $\delta{\hat x}$ of the extended coordinates. Then $\delta{\hat S} = 0$
gives the coordinate equations of motion:
\renewcommand{\theequation}{\thesection.\arabic{subsection}\alph{equation}}
\setcounter{subsection}{43}
\setcounter{equation}{0}
\begin{eqnarray}
\partial_m\left( \sum_w {\tilde h}_{(u+w)}^{mn} \partial_n {\hat x}^{n(r)\mu w}\right) = 0 \qquad \mbox{(bulk)} \label{eq2.43a} \\
\sum_{u,v} {\tilde h}_{(u+v)}^{m1} \partial_m {\hat x}^{n(r)\mu u} 
{\mathcal G}_{n(r)\mu;n(s)\nu} \delta {\hat x}^{n(s)\nu v} = 0 \quad\,\,\, \mbox{ at } \xi = 0,\pi. \label{eq2.43b}
\end{eqnarray}
The variational boundary conditions (2.43b) at the branes are in fact solved 
in the form
\renewcommand{\theequation}{\thesection.\arabic{equation}}
\setcounter{equation}{43}
\begin{equation}
\label{eq2.44}
\sum_{u,v} {\tilde h}_{(u+v)}^{m1} \partial_m {\hat x}^{n(r)\mu u} {\mathcal G}_{n(r)\mu;n(s)\nu} {\dot {\hat x}}^{n(s)\nu v} = 0\quad\,\,\, \mbox{ at } \xi = 0,\pi
\end{equation}
by the following coordinate-space boundary conditions:
\renewcommand{\theequation}{\thesection.\arabic{subsection}\alph{equation}}
\setcounter{subsection}{45}
\setcounter{equation}{0}
\begin{eqnarray}
\xi = 0: & &{\dot {\hat x}}^{n(r)\mu 0} =\, \partial_{\xi} {\hat x}^{n(r)\mu 1} = 0 \label{eq2.45a} \\
& &{\tilde h}_{(0)}^{01} =\, {\tilde h}_{(0)}^{11} = 0 \label{eq2.45b} \\
\xi = \pi: & &{\dot {\hat x}}^{n(r)\mu u} +\, i\tan (\tfrac {n(r)\pi}{\rho(\sigma)}) \sum_w {\tilde h}_{(w)}^{11} \partial_{\xi} {\hat x}^{n(r)\mu,u-w} = 0 \label{eq2.45c} \\
& &{\tilde h}_{(u)}^{01} = 0,\quad {\bar u} = 0,1. \label{eq2.45d}
\end{eqnarray}
These conditions are equivalent to those obtained earlier from phase 
space. For example the relation (2.45c) is nothing but 
the boundary condition (2.29b) written now with Eq. (2.41d) in terms of the 
extended inverse metric density. The condition (2.44) at $\pi$ is then
solved by (2.45c) because the identity
\renewcommand{\theequation}{\thesection.\arabic{equation}}
\setcounter{equation}{45}
\begin{equation}
\label{eq2.46}
\sum_{n(r),n(s)} \sum_{u,v,w} {\tilde h}_{(u+v+w)}^{11} {\tilde h}_{(w)}^{11} \tan (\tfrac {n(r)\pi}{\rho(\sigma)}) \partial_{\xi} {\hat x}^{n(r)\mu u} {\mathcal G}_{n(r)\mu;n(s)\nu} \partial_{\xi} {\hat x}^{n(s)\nu v} = 0
\end{equation}
holds by $n(r) \to -n(r)$ symmetry. For clarity I have here temporarily 
reinstated the implied sums over the spectral indices.

\setcounter{subsection}{6}
\subsection{Identification of the Extended Metric}
\label{sec2.7}

Following our discussion of the Polyakov gauge in Subsec. 2.4, it is 
clear that the extended inverse metric density ${\tilde h}_{(u)}^{mn}$ 
must be further decomposed to obtain the true {\it extended metric} ${\hat h}_{mn}^{(u)}$
of $\mathbb{Z}_2$-permutation gravity. The correct decomposition is
\renewcommand{\theequation}{\thesection.\arabic{subsection}\alph{equation}}
\setcounter{subsection}{47}
\setcounter{equation}{0}
\begin{equation}
{\tilde h}_{(u)}^{mn} = \sum_w {\hat h}_{(u+w)}^{mn} {\hat H}^{(w)},\quad\,\, {\bar u} = 0,1,\,\,\,\, m,n \in (0,1) \label{eq2.47a} 
\end{equation}
\begin{equation}
\sum_w {\hat h}_{(u+w)}^{mp} {\hat h}_{pn}^{(w+v)} = \delta_n^m \delta_{u-v,0 \!\!\!\!\mod 2} \label{eq2.47b} 
\end{equation}
\begin{equation}
{\hat H}^{(u)} \equiv \tfrac {1}{2} \left(\sqrt{-\det(\textstyle{\sum_w} {\hat h}_{mn}^{(w)})} + (-1)^u \sqrt{-\det( \textstyle{\sum_w} (-1)^w {\hat h}_{mn}^{(w)})}\right) \label{eq2.47c} 
\end{equation}
where ${\hat h}_{mn}^{(u)}$ and ${\hat h}_{(u)}^{mn}$ are respectively 
the  extended metric and its inverse. The determinant in Eq. (2.47c) is  $\det(A_{mn}) = A_{00}A_{11} - (A_{01})^2$.
Then it is straightforward to check that the extended, 
$\mathbb{Z}_2$-twisted diffeomorphisms
\renewcommand{\theequation}{\thesection.\arabic{subsection}\alph{equation}}
\setcounter{subsection}{48}
\setcounter{equation}{0}
\begin{equation}
\delta {\hat h}_{mn}^{(u)} = \sum_w ({\hat \beta}^{pw} \partial_p {\hat h}_{mn}^{(u-w)} + \partial_m {\hat \beta}^{pw} {\hat h}_{pn}^{(u-w)} + \partial_n {\hat \beta}^{pw} {\hat h}_{pn}^{(u-w)}) \label{eq2.48a} 
\end{equation}
\begin{equation}
\delta {\hat h}_{(u)}^{mn} = \sum_w ({\hat \beta}^{pw} \partial_p {\hat h}_{(u+w)}^{mn} - \partial_p {\hat \beta}^{mw} {\hat h}_{(u+w)}^{pn} - \partial_p {\hat \beta}^{nw} {\hat h}_{(u+w)}^{pm}) \label{eq2.48b} 
\end{equation}
\begin{equation}
\delta {\hat H}^{(u)} = \partial_m \left( \sum_w {\hat \beta}^{mw}{\hat H}^{(u-w)}\right),\,\,\, {\bar u} = 0,1 \label{eq2.48c}
\end{equation}
are consistent and reproduce the extended diffeomorphisms (2.38b) of the 
extended inverse metric density. Correspondingly, any object which 
transforms like ${\hat H}^{(u)}$ in (2.48c) will be called an extended 
scalar density.

Note that the extended metric
\renewcommand{\theequation}{\thesection.\arabic{equation}}
\setcounter{equation}{48}
\begin{equation}
\label{eq2.49}
{\hat h}_{mn}^{(u)} = \begin{pmatrix} {\hat h}_{00}^{(u)} & {\hat h}_{01}^{(u)} \\ {\hat h}_{10}^{(u)} & {\hat h}_{11}^{(u)} \end{pmatrix},\quad {\hat u} = 0,1
\end{equation}
has {\it six} independent degrees of freedom, while the extended inverse 
metric density has only four. This tells us that the extended metric 
contains {\it two} Weyl degrees of freedom. Indeed, I will argue in 
Subsec. 2.9 that an appropriate parametrization of the extended metric in 
the (completely gauge-fixed) conformal gauge is
\begin{equation}
\label{eq2.50}
{\hat h}_{mn}^{(u)} = \frac {1}{2} \eta_{mn} (e^{-{\hat \phi}_0} + (-1)^u 
e^{-{\hat \phi}_1})
\end{equation}
where $\{{\hat \phi}_I,\ I = 0,1\}$ are the extended Weyl degrees of 
freedom and $\eta_{mn}$ is the flat world-sheet metric. As a check on this 
form, one easily computes
\renewcommand{\theequation}{\thesection.\arabic{subsection}\alph{equation}}
\setcounter{subsection}{51}
\setcounter{equation}{0}
\begin{equation}
{\hat H}^{(u)} = \tfrac {1}{2} (e^{-{\hat \phi}_0} + (-1)^u e^{-{\hat \phi}_1}) \label{eq2.51a} 
\end{equation}
\begin{equation}
{\hat h}_{(u)}^{mn} = \tfrac {1}{2} \eta^{mn} (e^{{\hat \phi}_0} + (-1)^u e^{{\hat \phi}_1}) \label{eq2.51b} 
\end{equation}
\begin{equation}
\to {\tilde h}_{(u)}^{mn} = \eta^{mn} \delta_{u,0 \!\!\!\!\mod 2} \label{eq2.51c}
\end{equation}
as required in the conformal gauge.

Identification of the extended metric also allows us to consider the 
$\mathbb{Z}_2$-gravitational equations of motion, which are obtained by arbitrary
variation $\delta {\hat h}_{mn}^{(u)}$ of the final form of the extended action:
\renewcommand{\theequation}{\thesection.\arabic{equation}}
\setcounter{equation}{51}
\begin{equation}
\label{eq2.52}
{\hat S} = \frac {1}{4\pi} \int dt \int_0^{\pi} d\xi \sum_{u,v,w} {\hat h}_{(u+v+w)}^{mn} {\hat H}^{(u)} \partial_m 
{\hat x}^{n(r)\mu v} {\mathcal G}_{n(r)\mu;n(s)\nu} \partial_n {\hat x}^{n(s)\nu w}.
\end{equation}
Useful identities in this computation include the following
\renewcommand{\theequation}{\thesection.\arabic{subsection}\alph{equation}}
\setcounter{subsection}{53}
\setcounter{equation}{0}
\begin{equation}
\sum_w {\hat H}_{(u+w)}^{-1} {\hat H}^{(w+v)} = \delta_{u-v,0 \!\!\!\!\mod 2} \label{eq2.53a} 
\end{equation}
\begin{equation}
\delta {\hat H}^{(u)} = \tfrac {1}{2} \sum_{v,w} {\hat H}^{(u+v-w)} {\hat h}_{(v)}^{pq} \delta {\hat h}_{pq}^{(w)} \label{eq2.53b} 
\end{equation}
\begin{equation}
\delta {\hat h}_{(u)}^{mn} = -\sum_{w,v} {\hat h}_{(w)}^{mp} {\hat h}_{(v)}^{qn} \delta {\hat h}_{pq}^{(w+v-u)} \label{eq2.53c}
\end{equation}
where the quantity ${\hat H}^{-1}$ is defined by Eq. (2.53a). One then 
finds the extended $\mathbb{Z}_2$-gravitational stress tensor and 
equations of motion:
\renewcommand{\theequation}{\thesection.\arabic{subsection}\alph{equation}}
\setcounter{subsection}{54}
\setcounter{equation}{0}
\begin{equation}
{\hat \theta}_{mn}^{(u)} \equiv \sum_{xyv} {\hat H}_{(v)}^{-1} {\hat 
h}_{mp}^{(x)} {\hat h}_{nq}^{(y)} \left(\delta\hat S /\delta {\hat  h}_{pq}^{(x+y-v-u)}\right) \label{eq2.54a} 
\end{equation}
\begin{equation}
\hspace{.5in}= \tfrac {1}{2} \left({\hat {\mathcal L}}_{mn}^{(u)} - \tfrac {1}{2} \sum_{v,w} {\hat h}_{mn}^{(v)} {\hat h}_{(w)}^{pq}{\hat {\mathcal L}}_{pq}^{(w-v+u)}\right) \label{eq2.54b} 
\end{equation}
\begin{equation}
{\hat {\mathcal L}}_{mn}^{(u)} \equiv \tfrac {1}{4\pi} \sum_v \partial_m {\hat x}^{n(r)\mu v} {\mathcal G}_{n(r)\mu;n(s)\nu}\partial_n {\hat x}^{n(s)\nu,u-v} \label{eq2.54c} 
\end{equation}
\begin{equation}
\sum_v {\hat h}_{(u+v)}^{mn} {\hat \theta}_{mn}^{(v)} = 0 \label{eq2.54d} 
\end{equation}
\begin{equation}
{\hat \theta}_{mn}^{(u)} = 0,\quad {\bar u} = 0,1. \label{eq2.54e}
\end{equation}
The extended trace conditions in Eq. (2.54d) hold independent of the 
gravitational equations of motion in (2.54e).

In the conformal gauge, these results reduce to the following 
Weyl-field-independent forms
\renewcommand{\theequation}{\thesection.\arabic{subsection}\alph{equation}}
\setcounter{subsection}{55}
\setcounter{equation}{0}
\begin{equation}
\eta^{mn}{\hat \theta}_{mn}^{(u)} = 0 \label{eq2.55a} 
\end{equation}
\begin{equation}
{\hat \theta}_{mn}^{(u)} = \tfrac {1}{2} ({\hat {\mathcal L}}_{mn}^{(u)} - \tfrac {1}{2} \eta_{mn} \eta^{pq} {\hat {\mathcal L}}_{pq}^{(u)}) = 0,\quad {\bar u} = 0,1 \label{eq2.55b}
\end{equation}
after using the form (2.50) of the extended metric in the conformal 
gauge. As expected from the extended Hamiltonian description, these 
conditions are nothing but linear combinations of the classical {\it extended Virasoro constraints} of each 
twisted open-string CFT
\renewcommand{\theequation}{\thesection.\arabic{equation}}
\setcounter{equation}{55}
\begin{equation}
\label{eq2.56}
{\hat \theta}_u^{\pm}(\xi,t) = 0,\quad {\bar u} = 0,1.
\end{equation}
where the coordinate-space form of $\{{\hat \theta}_{u}^{\pm}\}$ was given in 
Eq. (2.36).

\setcounter{subsection}{7}
\subsection{Extended Nambu Action}
\label{sec2.8}

The quantity $\hat {\mathcal L}_{mn}^{(u)}$ of the previous subsection 
is constructed entirely from the extended coordinates and transforms 
under the $\mathbb{Z}_2$-twisted diffeomorphisms like the extended metric:
\renewcommand{\theequation}{\thesection.\arabic{subsection}\alph{equation}}
\setcounter{subsection}{57}
\setcounter{equation}{0}
\begin{equation}
{\hat {\mathcal L}}_{mn}^{(u)} \approx \sum_v \partial_m {\hat x}^{n(r)\mu v} {\mathcal G}_{n(r)\mu;u(s)\nu} \partial_n {\hat x}^{n(s)\nu,u-v} \label{eq2.57a} 
\end{equation}
\begin{equation}
\delta{\hat x}^{n(r)\mu u} = \sum_v {\hat \beta}^{mv} \partial_m {\hat x}^{n(r)\mu,u-v} \label{eq2.57b} 
\end{equation}
\begin{equation}
\delta {\hat {\mathcal L}}_{mn}^{(u)} = 
\sum_v ({\hat \beta}^{pv} \partial_p {\hat {\mathcal L}}_{mn}^{(u-v)} +
 \partial_m {\hat \beta}^{pv} {\hat {\mathcal L}}_{pn}^{(u-v)} +
  \partial_n {\hat \beta}^{pv} {\hat {\mathcal L}}_{pm}^{(u-v)}). \label{eq2.57c}
\end{equation}
This identifies  ${\mathcal L}_{mn}^{(u)}$ as the natural candidate for 
the extended form of the induced world-sheet metric.

We are then led to consider the following extended action of Nambu-type
\renewcommand{\theequation}{\thesection.\arabic{subsection}\alph{equation}}
\setcounter{subsection}{58}
\setcounter{equation}{0}
\begin{equation}
{\hat {\hat S}} \approx \int dt \int_0^{\pi} d\xi\, {\hat {\hat H}}^{(0)} \label{eq2.58a} 
\end{equation}
\begin{equation}
{\hat {\hat H}}^{(u)} \equiv \tfrac {1}{2} \left( 
\sqrt{-\det(\textstyle{\sum_v} {\hat {\mathcal L}}_{mn}^{(v)})} + (-1)^u \sqrt{-\det(\textstyle{\sum_v} (-1)^v {\hat {\mathcal L}}_{mn}^{(v)})} \right) \label{eq2.58b} 
\end{equation}
\begin{equation}
\delta {\hat {\hat H}}^{(0)} = \partial_p \left( \sum_u {\hat \beta}^{pu} {\hat {\hat H}}^{(-u)}\right) \label{eq2.58c}
\end{equation}
for each twisted open-string sector of $U(1)^d/({\mathbb Z}_2(w.s.)\times 
H)$. Eq. (2.58c) tells us in particular that the quantity ${\hat {\hat H}}^{(u)}$ is an 
extended scalar density and, using the boundary conditions (2.41) on the extended diffeomorphism 
parameters $\hat \beta$, one finds that this action is invariant under 
the extended diffeomorphisms when ${\hat {\hat H}}^{(1)}(0,t) = 0$.

For brevity, I will not pursue further analysis of these Nambu-like 
systems here. For future work, I note however that the four degrees of 
freedom of the extended diffeomorphisms should allow a ``light-cone 
gauge'' in which the number of independent (transverse) coordinate 
degrees of freedom is $2d-4$. See also my more detailed remarks on the 
extended Nambu actions of the permutation orbifolds in Subsec. 3.4 and my 
concluding remarks in Subsec. 4.1 on the corresponding operator theories at
critical dimension $d=26$.

\setcounter{subsection}{8}
\subsection{Finite Invariance Transformations}
\label{sec2.9}

I collect here some preliminary remarks on the finite symmetries of the 
extended action (2.52) of Polyakov-type, including the extended Weyl 
invariance and the extended diffeomorphism group of 
$\mathbb{Z}_2$-permutation gravity.

The central step in this discussion is to define what corresponds to the 
``fields with twisted boundary conditions'' [29,37] of each sector:
\renewcommand{\theequation}{\thesection.\arabic{subsection}\alph{equation}}
\setcounter{subsection}{59}
\setcounter{equation}{0}
\begin{equation}
{\hat h}_{mn}^I \equiv \sum_u (-1)^{Iu} {\hat h}_{mn}^{(u)},\quad I = 0,1 \label{eq2.59a} 
\end{equation}
\begin{equation}
{\hat h}_I^{mn} \equiv \sum_u (-1)^{Iu} {\hat h}_{(u)}^{mn},\quad {\hat h}_I^{mp} {\hat h}_{pn}^I = \delta_n^m \label{2.59b} 
\end{equation}
\begin{equation}
{\hat H}^I \equiv \sum_u (-1)^{Iu} {\hat H}^{(u)} = \sqrt{-\det({\hat h}_{mn}^I)} \label{eq2.59c} 
\end{equation}
\begin{equation}
{\hat x}^{n(r)\mu I} \equiv \sum_u (-1)^{Iu} {\hat x}^{n(r)\mu u},\quad {\hat x}^{n(r)\mu u} =
 \tfrac {1}{2} \sum_I (-1)^{uI} {\hat x}^{n(r)\mu I}. \label{eq2.59d}
\end{equation}
All these transformations are invertible, as shown explicitly for the 
twisted coordinates in Eq. (2.59d).

After some algebra, one finds that the new fields {\it diagonalize} the 
extended action and its diffeomorphisms
\renewcommand{\theequation}{\thesection.\arabic{subsection}\alph{equation}}
\setcounter{subsection}{60}
\setcounter{equation}{0}
\begin{equation}
{\hat S} = \sum_I \int dt_I \int_0^{\pi} d\xi_I\, {\hat H}^I {\hat h}_I^{mn} \partial_m {\hat x}^{n(r)\mu I} {\mathcal G}_{n(r)\mu;n(s)\nu} \partial_n {\hat x}^{n(s)\nu I} \label{eq2.60a} 
\end{equation}
\begin{equation}
\delta {\hat x}^{n(r)\mu I} = {\hat \beta}^{pI} \partial_p {\hat x}^{n(r)\mu I},\,\, \delta {\hat h}_{mn}^I = {\hat \beta}^{pI} \partial_p {\hat h}_{mn}^I + \partial_m {\hat \beta}^{pI} {\hat h}_{pn}^I + \partial_n {\hat \beta}^{pI} {\hat h}_{pm}^I \label{eq2.60b} 
\end{equation}
\begin{equation}
\beta^{pI} \equiv \sum_u (-1)^{Iu} {\hat \beta}^{pu}. \label{eq2.60c} 
\end{equation}
where I have relabelled the integration variables $t,\xi\rightarrow 
t_I,\xi_I$ separately in each term. In this form, one sees that the 
action is invariant -- at least locally in the bulk -- under the product of two 
diffeomorphism groups:
\renewcommand{\theequation}{\thesection.\arabic{subsection}\alph{equation}}
\setcounter{subsection}{61}
\setcounter{equation}{0}
\begin{equation}
\xi_I^{m\prime} =  \xi_I^{m\prime} (\{\xi_I^{p}\}),\quad m = 0,1,\,\, I = 0,1 \label{eq2.61a} 
\end{equation}
\begin{equation}
 {\hat x}^{n(r)\mu I\prime}(\{\xi_I^{p\prime}\}) = {\hat x}^{n(r)\mu I} (\{\xi_I^p\}) \label{eq2.61b} 
\end{equation}
\begin{equation}
{\hat h}_{mn}^{I\prime}(\{\xi_I^{p\prime}\}) = 
\tfrac {\partial \xi_I^r}{\partial \xi_I^{m\prime}} \tfrac {\partial\xi_I^s}{\partial\xi_I^{n\prime}}
 {\hat h}_{rs}^I(\{\xi_I^p\}). \label{eq2.61c}
\end{equation}
Up to coupling at the branes then (see below), each of the two metrics ${\hat h}_{mn}^I$ 
transforms as an ordinary metric under its group.

In principle, one can use various finite symmetries of the action (2.60a) 
 and invertibility of the definitions (2.59) to 
work out the form of these transformations in the original 
$u$-basis. Again up to coupling at the branes, consider the simple 
example of extended Weyl invariance:
\renewcommand{\theequation}{\thesection.\arabic{subsection}\alph{equation}}
\setcounter{subsection}{62}
\setcounter{equation}{0}
\begin{eqnarray}
{\hat h}_{mn}^I &\to &e^{-{\hat\sigma}_I} {\hat h}_{mn}^I: \label{eq2.62a}\\
{\hat h}_{mn}^{(u)} &\to &\tfrac {1}{2} \sum_w (e^{-{\hat \sigma}_0} + (-1)^{u+w} e^{-{\hat\sigma}_1}) {\hat h}_{mn}^{(w)} \label{eq2.62b} \\
{\tilde h}_{(u)}^{mn} &\to &{\tilde h}_{(u)}^{mn},\quad {\hat S}\,\, \to\,\,\, {\hat S}. \label{eq2.62c}
\end{eqnarray}
Similarly, the following steps 
\renewcommand{\theequation}{\thesection.\arabic{subsection}\alph{equation}}
\setcounter{subsection}{63}
\setcounter{equation}{0}
\begin{eqnarray}
h_{mn}^I &= &e^{-{\hat\phi}_I} \eta_{mn}: \label{eq2.63a} \\
{\hat h}_{mn}^{(u)} &= &\tfrac {1}{2} \eta_{mn} (e^{-{\hat \phi}_0} + (-1)^u e^{-{\hat \phi}_1}) \label{eq2.63b}
\end{eqnarray}
give the parametrization (2.51) quoted above for the extended metric in the conformal gauge.

I leave for another time and place the precise form of the finite 
$\mathbb{Z}_2$-diffeomorphisms in the $u$-basis. For this application in 
particular however, I emphasize that the decoupled form (2.60) of the 
bulk Lagrange density in the $I$-basis is quite {\it deceptive}, the 
complexity of the theory -- and in particular its fractional modeing --  being 
hidden in the unfamiliar boundary conditions  
\renewcommand{\theequation}{\thesection.\arabic{subsection}\alph{equation}}
\setcounter{subsection}{64}
\setcounter{equation}{0}
\begin{equation}
\sum_I (-1)^I {\tilde h}_I^{00}(0) =\, \sum_I (-1)^I {\tilde h}_I^{11}(0) =\, \sum_I {\tilde h}_I^{01}(0) = 0 \label{eq2.64a} 
\end{equation}
\begin{equation}
\sum_I {\dot {\hat x}}^{n(r)\mu I}(0) =\, \sum_I (-1)^I \partial_{\xi} {\hat x}^{n(r)\mu I}(0) = 0 \label{eq2.64b} 
\end{equation}
\begin{equation}
{\tilde h}_I^{01}(\pi) = 0 \label{eq2.64c} 
\end{equation}
\begin{equation}
{\dot {\hat x}}^{n(r)\mu I}(\pi) +\, i\tan (\tfrac {n(r)\pi}{\rho(\sigma)}) {\tilde h}_I^{11}(\pi) \partial_{\xi} {\hat x}^{n(r)\mu I}(\pi) = 0 \label{eq2.64d}
\end{equation}
which couple the $I$-basis fields at the branes, even in the conformal gauge. 

I finally mention that the basis change (2.59d) of the extended 
coordinates allows the extended action (2.58) of Nambu-type to be similarly
expressed in terms of two ordinary untwisted Nambu actions [65], but the coupled
coordinate boundary conditions at the branes are more intricate in this
case.

\setcounter{section}{2}
\section{General Permutation Gravity\\ \hspace*{.3in} in the Permutation Orbifolds}
\label{sec3}

\setcounter{subsection}{0}
\subsection{Extended Polyakov Hamiltonian}
\label{sec3.1}

The (closed-string) WZW permutation orbifolds [27,29,31-33,35,36]
\renewcommand{\theequation}{\thesection.\arabic{equation}}
\setcounter{equation}{0}
\begin{equation}
\label{eq3.1}
\frac {A_g(H)}{H},\quad g = \oplus_{I=0}^{K-1} {\mathbf{g}}^I,\,\,\,\,\, 
{\mathbf{g}}^I \simeq \mathbf{g},\,\,\,\,\, H = H(\mbox{perm}) \subset \mbox{Aut}(g)
\end{equation}
have now been studied in considerable detail, where $H(\mbox{perm})$ is 
any permutation group which acts on $K$ copies of $\mathbf{g}$ in $g$. We 
know in particular that the sectors $\{\sigma\}$ of these orbifolds are 
labelled by the conjugacy classes of $H(\mbox{perm})$,  
and each sector lives at central 
charge $\hat c= Kc_{\mathbf{g}}$ where $c_{\mathbf{g}}$ is the central 
charge of the affine-Sugawara construction [2,3,4,7,11] on $\mathbf{g}$. Moreover, each
twisted sector $(\sigma\neq 0)$ is governed by the following left-mover orbifold 
Virasoro algebra [35]
\renewcommand{\theequation}{\thesection.\arabic{subsection}\alph{equation}}
\setcounter{subsection}{2}
\setcounter{equation}{0}
\begin{eqnarray}
& &[{\hat L}_{{\hat \jmath} j}(m + \tfrac {{\hat \jmath}}{f_j(\sigma)}),\ {\hat L}_{{\hat l}l}(n + \tfrac {{\hat l}}{f_l(\sigma)})] \label{eq3.2a} \\
& &\qquad = \delta_{jl} \{(m-n+ \tfrac {{\hat \jmath} - {\hat l}}{f_j(\sigma)}) {\hat L}_{{\hat \jmath}+{\hat l},j}(m+n+ \tfrac {{\hat \jmath} + {\hat l}}{f_j(\sigma)}) \nonumber \\
& &\qquad + \tfrac {c_{\mathbf{g}}f_j(\sigma)}{12} (m + \tfrac {{\hat \jmath}}{f_j(\sigma)}) ((m + \tfrac {{\hat \jmath}}{f_j(\sigma)})^2 - 1)\, \delta_{m+u+ \tfrac {{\hat \jmath} + {\hat l}}{f_j(\sigma)},0} \nonumber\\ 
&&\qquad \bar{\hat \jmath},\bar{\hat l} = 0,1,\dots,f_j(\sigma)-1,\quad \sum_j f_j(\sigma) = K \label{eq3.2b}
\end{eqnarray}
as well as  a  commuting set of (rectified) right-mover 
copies $\{{\hat {\bar L}}_{{\hat \jmath}j}^{\#}\}$.  The physical  
Virasoro generators of each twisted sector  
\renewcommand{\theequation}{\thesection.\arabic{subsection}\alph{equation}}
\setcounter{subsection}{3}
\setcounter{equation}{0}
\begin{equation}
{\hat L}_{\sigma}(m) \equiv \sum_j {\hat L}_{0j}(m),\quad {\hat {\bar L}}_{\sigma}(m) \equiv \sum_j {\hat {\bar L}}_{0j}^{\#}(m) \label{eq3.3a}
\end{equation}
\begin{equation}
 {\hat c}\, =\, {\hat {\bar c}} \,=\, c\, =\, K c_{\mathbf{g}}  \label{eq3.3b}
\end{equation}
are twisted affine-Sugawara constructions with the same central charges (3.3b) as the untwisted sector $\sigma = 0$.
 
The extended Virasoro algebra in Eq. (3.2) is given in the standard [33-35,37] 
cycle notation for the corresponding representative element $h_{\sigma} 
\in H(\mbox{perm})$, where $j$ indexes the disjoint cycles of size 
$f_{j}(\sigma)$ and $\hat j$ indexes the position in the $j$th cycle. As 
examples, one has
\renewcommand{\theequation}{\thesection.\arabic{subsection}\alph{equation}}
\setcounter{subsection}{4}
\setcounter{equation}{0}
\begin{eqnarray}
& &{\mathbb Z}_{\lambda}: K = \lambda,\ f_j(\sigma) = \rho(\sigma),\ {\hat {\bar \jmath}} = 0,\dots,\rho(\sigma)-1, \label{eq3.4a} \\
& &\qquad  j = 0,\dots, \frac {1}{\rho(\sigma)} - 1,\ \sigma = 0,\dots,\rho(\sigma) - 1 \nonumber \\
& &{\mathbb Z}_{\lambda},\ \lambda \mbox{ prime}: \rho(\sigma) = \lambda,\ {\hat {\bar \jmath}} = 0,\dots,\lambda - 1, \label{eq3.4b} \\
& &\qquad j = 0,\ \sigma = 1,\,\dots,\lambda - 1 \nonumber \\
& &S_N: K = N,\ f_j(\sigma) = \sigma_j,\ \sigma_{j+1} \le \sigma_j, \label{eq3.4c} \\
& &\qquad j = 0,\dots,n(\vec{\sigma}) - 1,\ \sum_{j=0}^{n(\vec {\sigma})} \sigma_j = N. \nonumber
\end{eqnarray}
For the special case $H = {\mathbb Z}_2$, the extended Virasoro algebra 
(3.2) reduces to the order-two orbifold Virasoro algebra (2.2a) studied above 
for the orientation orbifolds. More generally, the extended algebra (3.2) is 
semisimple, with one component for each cycle $j$ in $h_{\sigma} \in H$.

For the classical development below, we will need the left- and 
right-mover extended stress tensors of each twisted sector [35]
\renewcommand{\theequation}{\thesection.\arabic{subsection}\alph{equation}}
\setcounter{subsection}{5}
\setcounter{equation}{0}
\begin{equation}
{\hat \theta}_{{\hat \jmath} j}(\xi,t) = \tfrac {1}{2\pi} \sum_{m \in {\mathbb Z}} {\hat L}_{{\hat \jmath} j} (m + \tfrac {\hat \jmath}{f_j(\sigma)}) e^{-i(m+\frac {\hat \jmath}{f_j(\sigma)})(t+\xi)} \label{eq3.5a} 
\end{equation}
\begin{equation}
{\hat {\bar \theta}}_{{\hat \jmath} j}(\xi,t) = \tfrac {1}{2\pi} \sum_{m \in {\mathbb Z}} {\hat {\bar L}}_{-{\hat \jmath},j}^{\#} (m - \tfrac {\hat \jmath}{f_j(\sigma)}) e^{-i(m - \frac {\hat \jmath}{f_j(\sigma)}(t-\xi)} \label{eq3.5b} 
\end{equation}
\begin{equation}
{\hat \theta}_{{\hat \jmath} j}(\xi + 2\pi,t) = e^{-2\pi i\frac {\hat \jmath}{f_j(\sigma)}} {\hat \theta}_{{\hat \jmath} j}(\xi,t),\,\, 
{\hat {\bar \theta}}_{{\hat \jmath} j}(\xi + 2\pi,t) = e^{-2\pi i\frac {\hat \jmath}{f_j(\sigma)}} {\hat {\bar \theta}}_{{\hat \jmath} j}(\xi,t) \label{eq3.5c}
\end{equation}
\begin{equation}
\{{\hat \theta}_{{\hat \jmath} j}(\xi,t),{\hat \theta}_{{\hat l}l}(\eta,t)\} = i\delta_{jl}(\partial_{\xi} - \partial_{\eta})({\hat \theta}_{{\hat \jmath}+{\hat l}}(\eta)\delta_{\frac {\hat \jmath}{f_j(\sigma)}}(\xi-\eta)) \label{eq3.5d} 
\end{equation}
\begin{equation}
\{{\hat {\bar \theta}}_{{\hat \jmath} j}(\xi,t),{\hat {\bar \theta}}_{{\hat l}l}(\eta,t)\} = -i \delta_{jl}(\partial_{\xi}-\partial_{\eta})({\hat {\bar \theta}}_{{\hat \jmath}+{\hat l}}(\eta)\delta_{\frac {\hat \jmath}{f_j(\sigma)}}(\xi-\eta)) \label{eq3.5e} 
\end{equation}
\begin{equation}
\{{\hat \theta}_{{\hat \jmath}j}(\xi,t),{\hat {\bar \theta}}_{{\hat l}l}(\eta,t)\} = 0 \label{eq3.5f}
\end{equation}
\begin{equation}
\delta_{\frac {\hat \jmath}{f_j(\sigma)}}(\xi-\eta) = \tfrac {1}{2\pi} \sum_{m \in {\mathbb Z}} e^{-i(m+\frac {\hat \jmath}{f_j(\sigma)})(\xi-\eta)} = \delta_{-\frac {\hat \jmath}{f_j(\sigma)}}(\eta-\xi). \label{eq3.5g}
\end{equation}
whose properties follow again from the classical analogue of Eq. (3.2b). The barred brackets here
follow from the unbarred because ${\hat {\bar 
\theta}}_{-{\hat \jmath},j}$ has the same brackets as ${\hat \theta}_{{\hat \jmath}j}$ with $\xi \to -\xi$.

Then following the development above for the orientation orbifolds, I 
define the (monodromy-invariant) extended Hamiltonian and gauge generator 
for each sector $\sigma$ of each permutation orbifold
\renewcommand{\theequation}{\thesection.\arabic{subsection}\alph{equation}}
\setcounter{subsection}{6}
\setcounter{equation}{0}
\begin{equation}
{\hat H}_{\sigma} \equiv \int_0^{\pi} d\xi \sum_j \sum_{{\hat \jmath} = 0}^{f_j(\sigma)-1} ({\hat v}^{{\hat \jmath}j} {\hat \theta}_{{\hat \jmath}j} + {\hat {\bar v}}^{{\hat \jmath}j}{\hat {\bar \theta}}_{{\hat \jmath}j}) \label{eq3.6a} 
\end{equation}
\begin{equation}
\equiv \int_0^{2\pi} d\xi ({\hat v}^{{\hat \jmath}j} {\hat \theta}_{{\hat \jmath}j} + {\hat {\bar v}}^{{\hat \jmath}j}{\hat {\bar \theta}}_{{\hat \jmath}j}) \label{eq3.6b} 
\end{equation}
\begin{equation}
{\hat G}_{\sigma} \equiv \int_0^{2\pi} d\xi ({\hat \epsilon}^{{\hat \jmath}j}{\hat \theta}_{{\hat \jmath}j} + {\hat {\bar \epsilon}}^{{\hat \jmath}j}{\hat {\bar \theta}}_{{\hat \jmath}j}) \label{eq3.6c} 
\end{equation}
\begin{equation}
{\hat {\mathcal O}}^{{\hat \jmath}j}(\xi+2\pi) =
 e^{2\pi i \frac {\hat \jmath}{f_j(\sigma)}} {\hat {\mathcal O}}^{{\hat \jmath}j}(\xi),\quad {\hat {\mathcal O}} = \{{\hat v},{\hat {\bar v}},{\hat \epsilon},{\hat {\bar \epsilon}}\} \label{eq3.6d} 
\end{equation}
\begin{equation}
{\dot {\hat A}} \equiv i\{{\hat H}_{\sigma},{\hat A}\},\quad\,\, \delta{\hat A} \equiv i\{{\hat G}_{\sigma},{\hat A}\} \label{eq3.6e} 
\end{equation}
\begin{equation}
\delta {\hat v}^{{\hat \jmath}j} \equiv \sum_{{\hat l}=0}^{f_j(\sigma)-1} {\hat \epsilon}^{{\hat \jmath}-{\hat l},j} \stackrel{\leftrightarrow}{\partial}_{\xi} {\hat v}^{{\hat l}j},\quad
 \delta{\hat{\bar v}}^{{\hat \jmath}j} \equiv \sum_{{\hat l}=0}^{f_j(\sigma)-1} {\hat {\bar v}}^{{\hat \jmath}-{\hat l},j} \stackrel{\leftrightarrow}{\partial}_{\xi} {\hat {\bar \epsilon}}^{{\hat l}j} \label{eq3.6f}
\end{equation}
where $\hat v,\hat{\bar v}$ are the multipliers and $\hat \epsilon, 
\hat{\bar\epsilon}$ are the (time-independent) extended gauge parameters. 
This gives in particular the properties of the gauge-variant stress 
tensors and multipliers
\renewcommand{\theequation}{\thesection.\arabic{subsection}\alph{equation}}
\setcounter{subsection}{7}
\setcounter{equation}{0}
\begin{equation}
\delta{\hat \theta}_{{\hat \jmath}j} = \sum_{{\hat l}=0}^{f_j(\sigma)-1} [\partial_{\xi}({\hat \theta}_{{\hat \jmath}+{\hat l},j} {\hat \epsilon}^{{\hat l}j}) + {\hat \theta}_{{\hat \jmath}+{\hat l},j} \partial_{\xi} {\hat \epsilon}^{{\hat l}j}] \label{eq3.7a} 
\end{equation}
\begin{equation}
\delta {\hat {\bar \theta}}_{{\hat \jmath}j} = -\sum_{{\hat l}=0}^{f_j(\sigma)-1} [\partial_{\xi}({\hat {\bar \theta}}_{{\hat \jmath}+{\hat l},j} {\hat {\bar \epsilon}}^{{\hat l}j}) + {\hat {\bar \theta}}_{{\hat \jmath}+{\hat l},j} \partial_{\xi} {\hat {\bar \epsilon}}^{{\hat l}j}] \label{eq3.7b} 
\end{equation}
\begin{equation}
{\dot {\hat \theta}}_{{\hat \jmath}j} = \sum_{{\hat l}=0}^{f_j(\sigma)-1} [\partial_{\xi} ({\hat\theta}_{{\hat \jmath}+{\hat l},j}{\hat v}^{{\hat l}j}) + {\hat \theta}_{{\hat \jmath}+{\hat l},j} \partial_{\xi} {\hat v}^{{\hat l}j}] \label{eq3.7c} 
\end{equation}
\begin{equation}
{\dot {\hat {\bar \theta}}}_{{\hat \jmath}j} = -\sum_{{\hat l}=0}^{f_j(\sigma)-1} [\partial_{\xi} ({\hat {\bar \theta}}_{{\hat \jmath}+{\hat l}} {\hat {\bar v}}^{{\hat l}j}) + {\hat {\bar \theta}}_{{\hat \jmath}+{\hat l},j} \partial_{\xi} {\hat {\bar v}}^{{\hat l}j}] \label{eq3.7d} 
\end{equation}
\begin{equation}
{\dot {\hat v}}^{{\hat \jmath}j} = -\sum_{{\hat l}=0}^{f_j(\sigma)-1} {\hat v}^{{\hat \jmath}-{\hat l},j} \stackrel{\leftrightarrow}{\partial}_{\xi} {\hat v}^{{\hat l}j},\quad {\dot {\hat {\bar v}}}^{{\hat \jmath}j} = \sum_{{\hat l}=0}^{f_j(\sigma)-1} {\hat {\bar v}}^{{\hat \jmath}-{\hat l},j} \stackrel{\leftrightarrow}{\partial}_{\xi} {\hat {\bar v}}^{{\hat l}j}. \label{eq3.7e}
\end{equation}
where (3.7e) follows from (3.7c), (3.7d) and the requirement that ${\dot 
{\hat H}}_{\sigma} = 0$. Then it is easily checked that the extended 
Hamiltonian is gauge-invariant  $\delta{\hat H}_{\sigma} = 0$ under the 
extended gauge group of the orbifold Virasoro algebra (3.2a). Note also 
that the monodromies, dynamics and gauge transformations in Eqs. (3.6), (3.7) do not mix the 
cycles $\{j\}$.

The corresponding multiplier equations of motion are the extended Virasoro 
(Polyakov) constraints
\renewcommand{\theequation}{\thesection.\arabic{subsection}\alph{equation}}
\setcounter{subsection}{8}
\setcounter{equation}{0}
\begin{eqnarray}
{\hat \theta}_{{\hat \jmath}j} &= &{\hat {\bar \theta}}_{{\hat \jmath}j}\, =\, 0,\quad {\bar {\hat \jmath}} = 0,1,\dots,f_j(\sigma)-1,\,\,\,\, \sum_j f_j(\sigma) = K \label{eq3.8a} \\
&\leftrightarrow &\{{\hat L}_{{\hat \jmath}j} (m + \tfrac {\hat \jmath}{f_j(\sigma)}) = {\hat {\bar L}}_{{\hat \jmath}j}^{\#} (m + \tfrac {\hat \jmath}{f_j(\sigma)}) = 0\} \label{eq3.8b}
\end{eqnarray}
so we will not be surprised to find these constraints again in the action 
formulation below. I emphasize that the number of extended Virasoro constraints 
(which is also the number of extended gauge degrees of freedom) is
\renewcommand{\theequation}{\thesection.\arabic{equation}}
\setcounter{equation}{8}
\begin{equation}
\label{eq3.9}
N_* = 2 \sum_j f_j(\sigma) = 2K.
\end{equation}
This counting holds in all sectors of each permutation orbifold, 
including the untwisted sector $\sigma = 0$ where $K$ is the number of 
copies of $\mathbf{g}$ in $g$.

Among possible Hamiltonian gauge choices, I mention the following: In the 
(completely-fixed) conformal gauge, the extended Hamiltonian reduces to 
the CFT Hamiltonian of each sector
\renewcommand{\theequation}{\thesection.\arabic{subsection}\alph{equation}}
\setcounter{subsection}{10}
\setcounter{equation}{0}
\begin{equation}
{\hat v}^{{\hat \jmath}j} = {\hat {\bar v}}^{{\hat \jmath}j} = \delta_{{\hat \jmath},0 \!\!\!\!\mod f_j(\sigma)} \label{eq3.10a}
\end{equation}
\begin{equation}
{\hat H}_{\sigma} = \int_0^{2\pi} d\xi \sum_j ({\hat \theta}_{0j} + {\hat {\bar \theta}}_{0j}) = {\hat L}_{\sigma}(0) + {\hat {\bar L}}_{\sigma}(0) \label{eq3.10b}
\end{equation}
where the extended stress tensors of each twisted closed-string CFT are given in Eq. (3.5). Other 
(partially-fixed) gauges of interest include the extended Polyakov gauge
\renewcommand{\theequation}{\thesection.\arabic{equation}}
\setcounter{equation}{10}
\begin{equation}
\label{eq3.11}
{\hat v}^{{\hat \jmath}j} = {\hat v}^j \delta_{{\hat \jmath},0 \!\!\!\!\mod f_j(\sigma)},\quad {\hat {\bar v}}^{{\hat \jmath}j} = {\hat {\bar v}}^j \delta_{{\hat \jmath},0 \!\!\!\!\mod f_j(\sigma)}
\end{equation}
which corresponds in fact to choosing a distinct (ordinary) Polyakov 
metric for each disjoint cycle $j$, and the Polyakov gauge
\renewcommand{\theequation}{\thesection.\arabic{subsection}\alph{equation}}
\setcounter{subsection}{12}
\setcounter{equation}{0}
\begin{equation}
{\hat v}^{{\hat \jmath}j} = {\hat v} \delta_{{\hat \jmath},0 \!\!\!\!\mod f_j(\sigma)},\quad {\hat {\bar v}}^{{\hat \jmath}j} = {\hat {\bar v}} \delta_{{\hat \jmath},0 \!\!\!\!\mod f_j(\sigma)} \label{eq3.12a} 
\end{equation}
\begin{equation}
{\hat H}_{\sigma} = \int_0^{2\pi} d\xi({\hat v}{\hat \theta} + {\hat {\bar v}}{\hat {\bar \theta}}) \label{eq3.12b} \end{equation}
\begin{equation}
{\hat \theta} \equiv \sum_j {\hat \theta}_{0j},\quad\,\, {\hat {\bar \theta}} \equiv \sum_j {\hat {\bar \theta}}_{0j} \label{eq3.12c}
\end{equation}
where ${\hat \theta}$ and ${\hat {\bar \theta}}$ are the physical stress 
tensors (see Eq. (3.3)) of sector $\sigma$. This is the ordinary Polyakov Hamiltonian of 
the sector, in which the same  Polyakov metric is chosen for all the cycles.

To go further, we need the explicit phase-space formulation of the 
extended stress tensors in each twisted sector of the permutation 
orbifolds \footnote{ Although WZW was used as an illustration above, the 
orbifold Virasoro algebra (3.2a) and the extended Hamiltonian system (3.6) hold for 
general permutation orbifolds, including sigma-model permutation 
orbifolds (see Ref. [37]).}. This data is given for the WZW permutation orbifolds in 
Refs. [35,37], but I limit the discussion here to the $(B = {\hat B} = 0)$ 
free-bosonic permutation orbifolds
\renewcommand{\theequation}{\thesection.\arabic{equation}}
\setcounter{equation}{12}
\begin{equation}
\label{eq3.13}
\frac {U(1)^{Kd}}{H(\mbox{perm})},\quad U(1)^{Kd} \equiv\, U(1)^d \times U(1)^d \times \dots (K \mbox{ times})
\end{equation}
all of whose sectors have central charge ${\hat c} = c = Kd$.

The free-bosonic permutation orbifolds have also been worked out in Refs. [33,35,37] as 
a special case on abelian $g$, and we may read off from these references 
(at $k$ = 1 for simplicity):
\renewcommand{\theequation}{\thesection.\arabic{subsection}\alph{equation}}
\setcounter{subsection}{14}
\setcounter{equation}{0}
\begin{equation}
{\hat \theta}_{{\hat \jmath}j} = \tfrac {1}{4\pi} \tfrac {G^{ab}}{f_j(\sigma)} \sum_{{\hat l}=0}^{f_j(\sigma)-1} {\hat J}_{{\hat l}aj} {\hat J}_{{\hat \jmath}-{\hat l},bj} \label{eq3.14a} 
\end{equation}
\begin{equation}
{\hat {\bar \theta}}_{{\hat \jmath}j} = \tfrac {1}{4\pi} \tfrac {G^{ab}}{f_j(\sigma)} \sum_{{\hat l}=0}^{f_j(\sigma)-1} {\hat {\bar J}}_{{\hat l}aj} {\hat J}_{{\hat \jmath}-{\hat l},bj} \label{eq3.14b} 
\end{equation}
\begin{equation}
{\hat J}_{{\hat \jmath}aj}(\xi + 2\pi) = e^{-2\pi i \frac {\hat \jmath}{f_j(\sigma)}} {\hat J}_{{\hat \jmath}aj}(\xi),\quad {\hat {\bar J}}_{{\hat \jmath}aj}(\xi + 2\pi)
= e^{-2\pi i \frac {\hat \jmath}{f_j(\sigma)}} {\hat {\bar J}}_{{\hat 
\jmath}aj}(\xi) \label{eq3.14c}
\end{equation}
\begin{eqnarray}
\{{\hat J}_{{\hat \jmath}aj}(\xi),{\hat J}_{{\hat l}bl}(\eta)\} &= &-\{{\hat {\bar J}}_{{\hat \jmath}aj}(\xi), {\hat {\bar J}}_{{\hat l}bl}(\eta)\} \label{eq3.14d} \\
&= &\delta_{jl} \delta_{{\hat \jmath}+{\hat l},0 \!\!\!\!\mod f_j(\sigma)} f_j(\sigma) G_{ab} \partial_{\xi} \delta_{\frac {\hat \jmath}{f_j(\sigma)}}(\xi-\eta) \label{eq3.14e} 
\end{eqnarray}
\begin{eqnarray}
{\hat J}_{{\hat \jmath}aj} &= &2\pi {\hat p}_{{\hat \jmath}aj} + \tfrac {f_j(\sigma)}{2} G_{ab} \partial_{\xi} {\hat x}^{-{\hat \jmath},bj} \label{eq3.14f} \\
{\hat {\bar J}}_{{\hat \jmath}aj} &= &-2\pi {\hat p}_{{\hat \jmath}aj} + \tfrac {f_j(\sigma)}{2} G_{ab} \partial_{\xi} {\hat x}^{-{\hat \jmath},bj} \label{eq3.14g} 
\end{eqnarray}
\begin{equation}
{\hat x}^{{\hat \jmath}aj}(\xi + 2\pi) = e^{2\pi i \frac {\hat \jmath}{f_j(\sigma)}} {\hat x}^{{\hat \jmath}aj},\quad {\hat p}_{{\hat \jmath}aj}(\xi + 2\pi) 
= e^{-2\pi i \frac {\hat \jmath}{f_j(\sigma)}} {\hat p}_{{\hat \jmath}aj}(\xi) \label{eq3.14h} 
\end{equation}
\begin{equation}
\{{\hat p}_{jaj}(\xi),{\hat x}^{{\hat l}bl}(\eta)\} = -i \delta_j^l \delta_a^b \delta_{{\hat \jmath}-{\hat l},0 \!\!\!\!\mod f_j(\sigma)} \delta_{\frac {\hat \jmath}{f_j(\sigma)}}(\xi-\eta). \label{eq3.14i}
\end{equation}
Here  $\{\hat x\}$ and $\{\hat p\}$ are the twisted coordinates and momenta 
(there are $Kd$ of each) in twisted sector $\sigma$ of each free-bosonic permutation 
orbifold, and each twisted current algebra in (3.14d,e) is called an abelian 
orbifold affine algebra [27].

In further detail, the quantity $G_{ab}$ (and its inverse $G^{ab}$) in 
(3.14) is the untwisted tangent-space metric for each closed string copy 
$U(1)^d$ in the symmetric sector $U(1)^{Kd}$, where the untwisted current 
algebras, permutations and H-eigenvalue problem read:
\renewcommand{\theequation}{\thesection.\arabic{subsection}\alph{equation}}
\setcounter{subsection}{15}
\setcounter{equation}{0}
\begin{equation}
[J_{aI}(m),J_{bJ}(n)] = [{\bar J}_{aI}(m),{\bar J}_{bJ}(n)] = m\delta_{IJ}G_{ab}\delta_{m+n,0} \label{eq3.15a} 
\end{equation}
\begin{equation}
J_{aI}(m)' = \omega(\sigma)_I{}^J J_{aJ}(m),\quad {\hat {\bar J}}_{aI}(m)'   
= \omega(\sigma)_I{}^J {\bar J}_{aJ}(m) \label{eqn3.15b}
\end{equation}
\begin{equation}
 \omega(\sigma) \in H(\mbox{perm})  \label{eq3.15c}
\end{equation}
\begin{equation}
\omega(\sigma)_I{}^J U^{\dag}(\sigma)_J{}^{{\hat \jmath}j} = U^{\dag}(\sigma)_I{}^{{\hat \jmath}j} e^{-2\pi i \frac {\hat \jmath}{f_j(\sigma)}},  \label{eq3.15d} 
\end{equation}
\begin{equation}
{\bar {\hat \jmath}} = 0,\dots,f_j(\sigma)-1,\quad \sum_j f_j(\sigma) = K \label{eq3.15e} 
\end{equation}
\begin{equation}
m,n \in {\mathbb Z},\,\,\, a,b = 1,\dots,d,\,\,\, I,J = 0,\dots,K-1. \label{eq3.15f}
\end{equation}
This is the same $G_{ab}$ introduced for the untwisted open string in 
Sec. 2, and the same value
\renewcommand{\theequation}{\thesection.\arabic{subsection}\alph{equation}}
\setcounter{subsection}{16}
\setcounter{equation}{0}
\begin{equation}
\frac {U(1)^{26K}}{H(\mbox{perm})} :\quad G_{ab} = G^{ab} = \begin{pmatrix} -1 & 0 \\ 0 & 1\!\!1 \end{pmatrix},\,\,\, a,b = 0,1,\dots,25 \label{eq3.16a}
\end{equation}
\begin{equation} 
{\hat c} = c = 26K \label{eq3.16b}
\end{equation}
can be chosen at any point in the discussion below to obtain the results for the critical 
permutation orbifolds in Minkowski space.

It is now straightforward to work out the Hamiltonian equation of motion 
of the twisted coordinates
\renewcommand{\theequation}{\thesection.\arabic{equation}}
\setcounter{equation}{16}
\begin{equation}
\label{eq3.17}
{\dot {\hat x}}^{{\hat \jmath}aj} = \tfrac {G^{ab}}{f_j(\sigma)} \sum_{{\hat l}=0}^{f_j(\sigma)-1} ({\hat v}^{{\hat l}j} {\hat J}_{{\hat l}-{\hat \jmath},bj} - {\hat {\bar v}}^{{\hat l}j} {\hat {\bar J}}_{{\hat l}-{\hat \jmath},bj})
\end{equation}
and use the phase-space realization (3.14f,g) of the currents to construct the 
extended action
\renewcommand{\theequation}{\thesection.\arabic{equation}}
\setcounter{equation}{17}
\begin{equation}
\label{eq3.18}
{\hat S}_{\sigma} = \int dt \int_0^{2\pi} d\xi \sum_{j,{\hat \jmath}} ({\dot {\hat x}}^{{\hat \jmath}aj} {\hat p}_{{\hat \jmath}aj} - {\hat v}^{{\hat \jmath}j} {\hat \theta}_{{\hat \jmath}j} - {\hat {\bar v}}^{{\hat \jmath}j} {\hat {\bar \theta}}_{{\hat \jmath}j})
\end{equation}
in each sector $\sigma$ of all the permutation orbifolds. Again following Refs. [63,61], we know that each extended action
is invariant $\delta {\hat S}_{\sigma} = 0$ under the following full 
time-dependent gauge transformations
\renewcommand{\theequation}{\thesection.\arabic{subsection}\alph{equation}}
\setcounter{subsection}{19}
\setcounter{equation}{0}
\begin{eqnarray}
\delta {\hat v}^{{\hat \jmath}j} &= &{\dot {\hat \epsilon}}^{{\hat \jmath}j} + \sum_{{\hat l}=0}^{f_j(\sigma)-1} {\hat \epsilon}^{{\hat \jmath}-{\hat l},j} \stackrel{\leftrightarrow}{\partial}_{\xi} {\hat v}^{{\hat l}j} \label{eq3.19a} \\
\delta {\hat {\bar v}}^{{\hat \jmath}j} &= &{\dot {\hat {\bar \epsilon}}}^{{\hat \jmath}j} + \sum_{{\hat l}=0}^{f_j(\sigma)-1} {\hat {\bar v}}^{{\hat \jmath}-{\hat l},j} \stackrel{\leftrightarrow}{\partial}_{\xi} {\hat {\bar \epsilon}}^{{\hat l}j} \label{eq3.19b}
\end{eqnarray}
and $\delta{\hat A}$ in Eq. (3.6e) for the matter fields. These 
transformations define the phase-space form of the extended, twisted 
diffeomorphisms of general permutation gravity.

\setcounter{subsection}{1}
\subsection{Extended Polyakov Actions}
\label{sec3.2}

In order to keep track of the branes, I followed the transition to 
coordinate space carefully for the open-string sectors of the orientation orbifolds.
Such detail is unnecessary for the closed-string sectors of the 
permutation orbifolds because boundary conditions are now replaced by {\it monodromies},
and these are simple to track for all fields. Indeed, consulting Eqs. 
(3.5c), (3.6d) and (3.14h), we see a $+$ or $-$ phase under 
$\xi\rightarrow \xi+2\pi$ for each up or down index $\hat\jmath$ respectively.

I therefore present only the final coordinate-space form of ${\hat 
S}_{\sigma}$ in Eq. (3.18), which I will call the {\it general extended action} 
of Polyakov-type:
\renewcommand{\theequation}{\thesection.\arabic{subsection}\alph{equation}}
\setcounter{subsection}{20}
\setcounter{equation}{0}
\begin{equation}
{\hat {\mathcal L}}_{{\hat \jmath}j}^{\sigma} \equiv \tfrac {1}{8\pi} \sum_{{\hat l},{\hat m} = 0}^{f_j(\sigma)-1} {\tilde h}_{({\hat \jmath}+{\hat l}+{\hat m})j}^{mn} \partial_m {\hat x}^{{\hat l}aj} (f_j(\sigma)G_{ab}) \partial_n {\hat x}^{{\hat m}bj} \label{eq3.20a}
\end{equation}
\begin{eqnarray} 
{\hat S}_{\sigma} &= &\int dt \int_0^{2\pi} d\xi \,\sum_j {\hat {\mathcal L}}_{0j}^{\sigma} \label{eq3.20b} \\
&= &\tfrac {1}{8\pi} \int dt \int_0^{2\pi} d\xi \,\sum_{j{\hat l}{\hat m}} {\tilde h}_{({\hat l}+{\hat m})j}^{mn} \partial_m {\hat x}^{{\hat l}aj} f_j(\sigma)G_{ab}\partial_n {\hat x}^{{\hat m}bj} \label{eq3.20c}
\end{eqnarray}
\begin{equation}
{\hat x}^{{\hat \jmath}aj}(\xi + 2\pi,t) = e^{2\pi i \frac {\hat 
\jmath}{f_j(\sigma)}} {\hat x}^{{\hat \jmath}aj} (\xi,t)  \label{eq3.20d}
\end{equation}
\begin{equation}
{\tilde h}_{({\hat \jmath})j}^{mn} (\xi + 2\pi,t) = e^{-2\pi i \frac {\hat \jmath}{f_j(\sigma)}} {\tilde h}_{({\hat \jmath})j}^{mn} (\xi,t) \label{eq3.20e}
\end{equation}
\begin{equation}
{\hat {\mathcal L}}_{{\hat \jmath}j}^{\sigma}(\xi + 2\pi, t) = 
 e^{-2\pi i \frac {\hat \jmath}{f_j(\sigma)}} {\hat {\mathcal L}}_{{\hat 
 \jmath}j}^{\sigma}(\xi, t). \label{eq3.20e}
\end{equation}
For each sector $\sigma$ of each permutation orbifold the general 
extended action density is cycle-separable and monodromy-invariant. The general 
permutation-twisted gravitational structure
 \renewcommand{\theequation}{\thesection.\arabic{subsection}\alph{equation}}
\setcounter{subsection}{21}
\setcounter{equation}{0}
\begin{equation}
{\tilde h}_{({\hat \jmath})j}^{mn} (\xi,t) = {\tilde h}_{({\hat 
\jmath})j}^{nm}(\xi,t) \label{eq3.21a}
\end{equation}
\begin{equation}
\hspace{.7in}\ m,n \in (0,1),\quad {\bar {\hat \jmath}} = 0, \dots,f_j(\sigma)-1,\quad 
 \sum_j f_j(\sigma) = K \label{eq3.21b}
\end{equation}
collects all dependence on the phase-space multipliers $\{{\hat v},{\hat 
{\bar v}}\}$, and hence possesses $N_* = 2K$ independent degrees of 
freedom. I will again call this structure the {\it inverse extended 
metric density} of general permutation gravity, though we shall see 
momentarily that it is in fact a  set of inverse extended metric 
densities, one for each cycle $j$ of each representative element $h_{\sigma} \in H(\mbox{perm})$.

The coordinate-space form of the infinitesimal twisted diffeomorphisms (3.6e),(3.19a) 
of general permutation gravity is
\renewcommand{\theequation}{\thesection.\arabic{subsection}\alph{equation}}
\setcounter{subsection}{22}
\setcounter{equation}{0}
\begin{equation}
\delta{\hat x}^{{\hat \jmath}aj} = \sum_{{\hat l}=0}^{f_j(\sigma)-1} {\hat \beta}^{p{\hat l}j} \partial_p {\hat x}^{{\hat \jmath}-{\hat l},aj} \label{eq3.22a} 
\end{equation}
\begin{equation}
{\hat \beta}^{m{\hat \jmath}j}(\xi + 2\pi,t) = e^{2\pi i\frac {\hat \jmath}{f_j(\sigma)}} {\hat \beta}^{m{\hat \jmath}j}(\xi,t) \label{eq3.22b} 
\end{equation}
\begin{equation}
\delta{\tilde h}_{({\hat \jmath})j}^{mn} = \sum_{{\hat l}=0}^{f_j(\sigma)-1} (\partial_p({\hat \beta}^{p{\hat l}j} {\tilde h}_{({\hat \jmath}+{\hat l})j}^{mn}) - (\partial_p {\hat \beta}^{m{\hat l}j}) {\tilde h}_{({\hat \jmath}+{\hat l})j}^{pn} - (\partial_p {\hat \beta}^{n{\hat l}j} {\tilde h}_{({\hat \jmath}+{\hat l})j}^{pm})) \label{eq3.22c} 
\end{equation}
\begin{equation}
\delta {\hat {\mathcal L}}_{0j}^{\sigma} = \partial_p \left( \sum_{{\hat j}=0}^{f_j(\sigma)-1} {\hat \beta}^{p{\hat \jmath}j} {\hat {\mathcal L}}_{{\hat \jmath}j}^{\sigma}\right) \label{eq3.22d}
\end{equation}
where $\{\hat{\beta}\}$ are the $N_* = 2K$ independent coordinate-space 
extended diffeomorphism parameters. It follows that $\delta {\hat 
S}_{\sigma} = 0$, so that the general extended action has the expected 
invariance under the extended diffeomorphisms. Moreover, we see that the 
extended diffeomorphisms do not mix the different cycle-components $\{j\}$ of 
the extended inverse metric density.

I close this subsection with the twisted-coordinate equations of motion
\renewcommand{\theequation}{\thesection.\arabic{subsection}\alph{equation}}
\setcounter{subsection}{23}
\setcounter{equation}{0}
\begin{equation}
\partial_m \left( \sum_{{\hat l}=0}^{f_j(\sigma)-1} {\tilde h}_{({\hat \jmath}+{\hat l})j}^{mn} \partial_n {\hat x}^{{\hat l}aj}\right) = 0 \label{eq3.23a} 
\end{equation}
\begin{equation}
{\bar {\hat \jmath}} = 0,1,\dots,f_j(\sigma)-1,\quad \sum_j f_j(\sigma) = K,\quad a = 1,\dots,d \label{eq3.23b}
\end{equation}
which follow by arbitrary variation $\delta\hat x$ of the general 
extended action, and two simple checks of our results up to this point:
First, for the single twisted sector of the $\mathbb{Z}_2$-permutation 
orbifold, one sees that our results reduce locally to the {\it same} 
$\mathbb{Z}_2$-permutation gravity found in the (open-string) orientation 
orbifold sectors. In particular, the extended inverse metric density in 
this case takes the form ${\tilde h}_{({\hat \jmath})0}^{mn}$, ${\bar {\hat \jmath}} = 0,1$
because the non-trivial element of $\mathbb{Z}_2$ is a single cycle. This 
makes sense because the local gravitational structure is governed by an order-two orbifold Virasoro
algebra in both cases. Second, the general extended action (3.20) reduces in 
the case of the (completely-fixed) conformal gauge (3.10) to the known conformal-field-theoretic
action of each sector of the free-bosonic permutation orbifolds [37]
\renewcommand{\theequation}{\thesection.\arabic{subsection}\alph{equation}}
\setcounter{subsection}{24}
\setcounter{equation}{0}
\begin{equation}
{\tilde h}_{({\hat \jmath})j}^{mn} = \eta^{mn} \delta_{{\hat \jmath},0 \!\!\!\!\mod f_j(\sigma)},\quad \eta = \begin{pmatrix} 1 & 0 \\ 0 & -1 \end{pmatrix} \label{eq3.24a}
\end{equation}
\begin{equation}
{\hat S}_{\sigma} = \tfrac {1}{8\pi} \int dt \int_0^{2\pi} d\xi \eta^{mn} \sum_j f_j(\sigma) \sum_{{\hat \jmath}=0}^{f_j(\sigma)-1} \partial_m {\hat x}^{{\hat \jmath}aj} G_{ab} \partial_n {\hat x}^{-{\hat \jmath},bj} \label{eq3.24b}
\end{equation}
where $\eta$ is the flat-world sheet metric. The coordinate-monodromies 
given in Eq. (3.20d) are of course independent of gauge choice,
while the monodromy (3.20e)  of the inverse extended metric density 
is now trivial because its conformal-gauge support is only at $\bar{\hat{\jmath}} 
= 0$.

\setcounter{subsection}{2}
\subsection{The Twisted Permutation Gravities}
\label{sec3.3}

I turn next to the identification of the {\it extended world-sheet metric} of 
each twisted permutation gravity:
\renewcommand{\theequation}{\thesection.\arabic{equation}}
\setcounter{equation}{24}
\begin{equation}
\label{eq3.25}
{\hat h}_{mn}^{({\hat \jmath})j} = {\hat h}_{nm}^{({\hat \jmath})j},\quad\,\, {\bar {\hat \jmath}} = 0,\dots,f_j(\sigma)-1,\,\,\, \sum_j f_j(\sigma) = K,\,\,\, m,n \in (0,1).
\end{equation}
Generalizing our discussion of $\mathbb{Z}_2$-permutation gravity in Sec. 
2, the extended metric (and its inverse ${\hat h}_{({\hat 
\jmath})j}^{mn}$) can be identified by the following decomposition of the extended
inverse metric density:
\renewcommand{\theequation}{\thesection.\arabic{subsection}\alph{equation}}
\setcounter{subsection}{26}
\setcounter{equation}{0}
\begin{equation}
{\tilde h}_{({\hat \jmath})j}^{mn} = \sum_{{\hat l} = 0}^{f_j(\sigma)-1} {\hat h}_{({\hat \jmath}+{\hat l})}^{mn} {\hat H}^{({\hat l})j},\quad\,\, {\hat h}_{({\hat \jmath})j}^{mn} = {\hat h}_{({\hat \jmath})j}^{nm} \label{eq3.26a} 
\end{equation}
\begin{equation}
\sum_{{\hat k}=0}^{f_j(\sigma)-1} {\hat h}_{({\hat \jmath}+{\hat k})j}^{mp} {\hat h}_{pn}^{({\hat k}+{\hat l})j} = \delta_n^m \delta_{{\hat \jmath}-{\hat l},0 \!\!\!\!\mod f_j(\sigma)},\,\, \forall\ j \label{eq3.26b} 
\end{equation}
\begin{equation}
{\hat H}^{({\hat \jmath})j} = \tfrac {1}{f_j(\sigma)} \sum_{J=0}^{f_j(\sigma)-1} e^{-2\pi i \frac {{\hat \jmath}J}{f_j(\sigma)}} \left(-\det\left( \sum_{{\hat l}=0}^{f_j(\sigma)-1} e^{2\pi i \frac {J{\hat l}}{f_j(\sigma)}} {\hat h}_{mn}^{({\hat l})j}\right)\right)^{\frac {1}{2}}. \label{eq3.26c}
\end{equation}
As above, the determinant in Eq. (3.26c) operates only on the $2\times2$ 
world-sheet indices $m,n$. It is then straightforward to check that the 
following permutation-twisted diffeomorphisms and monodromies
\renewcommand{\theequation}{\thesection.\arabic{subsection}\alph{equation}}
\setcounter{subsection}{27}
\setcounter{equation}{0}
\begin{equation}
\delta {\hat h}_{mn}^{({\hat \jmath})j} = \sum_{{\hat l}=0}^{f_j(\sigma)-1} ({\hat \beta}^{p{\hat l}j} \partial_p {\hat h}_{mn}^{({\hat \jmath}-{\hat l})j} + \partial_m {\hat \beta}^{p{\hat l}j} {\hat h}_{pn}^{({\hat \jmath}-{\hat l})j} + \partial_n {\hat \beta}^{p{\hat l}j} {\hat h}_{pm}^{({\hat \jmath}-{\hat l})j}) \label{eq3.27a} 
\end{equation}
\begin{equation}
\delta {\hat h}_{({\hat \jmath})j}^{mn} = \sum_{{\hat l}=0}^{f_j(\sigma)-1} ({\hat \beta}^{p{\hat l}j} \partial_p {\hat h}_{({\hat \jmath}+{\hat l})j}^{mn} - \partial_p{\hat \beta}^{m{\hat l}j} {\hat h}_{({\hat \jmath}+{\hat l})j}^{pn} - \partial_p {\hat \beta}^{n{\hat l}j} {\hat h}_{({\hat \jmath}+{\hat l})j}^{pm}) \label{eq3.27b} 
\end{equation}
\begin{equation}
\delta {\hat H}^{({\hat \jmath})j} = \partial_p \left( \sum_{{\hat l}=0}^{f_j(\sigma)-1} \beta^{p{\hat l}j} {\hat H}^{({\hat \jmath}-{\hat l})j}\right) \label{eq3.27c} 
\end{equation}
\begin{equation}
{\hat h}_{mn}^{({\hat \jmath})j} (\xi+2\pi,t) = e^{2\pi i \frac {\hat \jmath}{f_j(\sigma)}} {\hat h}_{mn}^{({\hat \jmath})j} (\xi,t) \label{eq3.27d} 
\end{equation}
\begin{equation}
{\hat h}_{({\hat \jmath})j}^{mn}(\xi+2\pi,t) = e^{-2\pi i \frac {\hat \jmath}{f_j(\sigma)}} {\hat h}_{({\hat \jmath})j}^{mn}(\xi,t) \label{eq3.27e} 
\end{equation}
\begin{equation}
{\hat H}^{({\hat \jmath})j}(\xi+2\pi,t) = e^{2\pi i \frac {\hat \jmath}{f_j(\sigma)}} {\hat H}^{({\hat \jmath})j}(\xi,t) \label{eq3.27f}
\end{equation}
are consistent and reproduce the corresponding transformations in (3.20e),\\ 
(3.22c) of the extended inverse metric density. We may take the 
transformations (3.27c),(3.27f) as defining a set of extended scalar 
densities, one for each cycle $j$ of $h_{\sigma}$, although only the 
$\bar{\hat{\jmath}} =0$ component of each has trivial monodromy. For the case 
$H(\mbox{perm}) = \mathbb{Z}_2$ the decomposition (3.26) and the twisted 
diffeomorphisms in (3.27) reduce to the $\mathbb{Z}_2$-gravitational results of 
Subsec. 2.7.

More generally, these results describe a distinct, twisted permutation 
gravity in each twisted sector of each permutation orbifold. 
Correspondingly, the permutation gravities are in $1-1$ correspondence 
with the conjugacy classes of any permutation group $H(\mbox{perm})$. Although
I have worked out the invariant actions only for the free-bosonic orbifolds
\renewcommand{\theequation}{\thesection.\arabic{equation}}
\setcounter{equation}{27}
\begin{equation}
\label{eq3.28}
{\hat S}_{\sigma} = \frac {1}{8\pi} \int dt \int_0^{2\pi} d\xi \sum_j \sum_{{\hat \jmath},{\hat l},{\hat k}=0}^{f_j(\sigma)-1} {\hat h}_{({\hat \jmath}+{\hat l}+{\hat k})j}^{mn} {\hat H}^{({\hat \jmath})j} \partial_m {\hat x}^{{\hat l}aj} f_j(\sigma)G_{ab}\partial_n {\hat x}^{{\hat k}bj} 
\end{equation}
the extended Hamiltonian formulation of Subsec. 3.1 tells us that the 
{\it same} $P$-gravitational structures will also appear in the extended actions 
of permutation orbifolds of general WZW and sigma models.

Note that the extended world-sheet metric ${\hat h}_{mn}^{({\hat 
\jmath})j}$ of each twisted sector $\sigma$ has
\renewcommand{\theequation}{\thesection.\arabic{equation}}
\setcounter{equation}{28}
\begin{equation}
\label{eq3.29}
N'_* = 3K
\end{equation}
independent components while there are only $N_* = 2K$ extended gauge 
degrees of freedom in each permutation gravity. Then the extended metric 
includes 
\renewcommand{\theequation}{\thesection.\arabic{equation}}
\setcounter{equation}{29}
\begin{equation}
\label{eq3.30}
N'_* - N_* = K
\end{equation}
twisted Weyl degrees of freedom, which cannot be gauged away and which do not
appear in the extended inverse metric density ${\tilde h}_{({\hat \jmath})j}^{mn}$.
This counting holds in all sectors of the permutation orbifolds, 
including the untwisted sector $\sigma = 0$ where we have $K$ copies of 
the ordinary Polyakov metric in the world-sheet description of $U(1)^{Kd}$.
Indeed, I will argue in the following subsection that the extended metric 
of twisted sector $\sigma$ can be parametrized in the (completely-fixed) conformal
gauge as
\renewcommand{\theequation}{\thesection.\arabic{subsection}\alph{equation}}
\setcounter{subsection}{31}
\setcounter{equation}{0}
\begin{equation}
{\hat h}_{mn}^{({\hat \jmath})j} = \eta_{mn} {\hat H}^{({\hat \jmath})j} \label{eq3.31a} 
\end{equation}
\begin{equation}
{\hat H}^{({\hat \jmath})j} =
 \tfrac {1}{f_j(\sigma)}
  \sum_{J=0}^{f_j(\sigma)-1} \exp\left( -2\pi i \tfrac {{\hat \jmath}J}{f_j(\sigma)} - \sum_{{\hat l}=0}^{f_j(\sigma)-1} e^{-2\pi i \frac {J{\hat l}}{f_j(\sigma)}} {\hat \phi}_{({\hat \jmath})j}\right) \label{eq3.31b} 
\end{equation}
\begin{equation}
{\hat h}_{({\hat \jmath})j}^{mn} = \eta^{mn}
 \tfrac {1}{f_j(\sigma)} 
\sum_{J=0}^{f_j(\sigma)-1} \exp\left( 2\pi i \tfrac {{\hat \jmath}J}{f_j(\sigma)} + \sum_{{\hat l}=0}^{f_j(\sigma)-1} e^{-2\pi  i \frac {J{\hat l}}{f_j(\sigma)}} {\hat \phi}_{({\hat \jmath})j} \right) \label{eq3.31c} 
\end{equation}
\begin{equation}
{\tilde h}_{({\hat \jmath})j}^{mn} = \eta^{mn} \delta_{{\hat \jmath},0 \!\!\!\!\mod f_j(\sigma)} \label{eq3.31d} 
\end{equation}
\begin{equation}
{\hat \phi}_{({\hat \jmath})j}(\xi + 2\pi,t) = e^{-2\pi i \frac {\hat \jmath}{f_j(\sigma)}} {\hat \phi}_{({\hat \jmath})j} (\xi,t), \label{eq3.31e} 
\end{equation}
\begin{equation}
{\bar {\hat \jmath}} = 0,\dots,f_j(\sigma)-1,\ \sum_j f_j(\sigma) = K \nonumber
\end{equation}
where $\eta$ is the flat world-sheet metric and $\{\hat\phi\}$ are the $K$ twisted Weyl fields.

As a final topic in this subsection, I consider the $P$-gravitational 
equations of motion, which are obtained by arbitrary variation $\delta{\hat h}_{mn}^{({\hat \jmath})j}$
of the extended metric in the extended action (3.28). After some algebra, 
the result is
\renewcommand{\theequation}{\thesection.\arabic{subsection}\alph{equation}}
\setcounter{subsection}{32}
\setcounter{equation}{0}
\begin{equation}
{\hat \theta}_{mn}^{{\hat \jmath}j} \equiv \tfrac {1}{2} \left( {\hat {\mathcal L}}_{mn}^{{\hat \jmath}j} - \tfrac {1}{2} \sum_{{\hat l},{\hat m}=0}^{f_j(\sigma)-1} {\hat h}_{mn}^{({\hat l})j} {\hat h}_{({\hat m})j}^{pq} {\hat {\mathcal L}}_{pq}^{{\hat m}-{\hat l}-{\hat \jmath},j}\right) \label{eq3.32a} 
\end{equation}
\begin{equation}
{\hat {\mathcal L}}_{mn}^{{\hat \jmath}j} \equiv \tfrac {1}{8\pi} \sum_{{\hat l}=0}^{f_j(\sigma)-1} \partial_m {\hat x}^{{\hat l}aj} f_j(\sigma) G_{ab} \partial_n {\hat x}^{{\hat \jmath}-{\hat l},bj} \label{eq3.32b} 
\end{equation}
\begin{equation}
\sum_{{\hat l}=0}^{f_j(\sigma)-1} {\hat h}_{({\hat \jmath}+{\hat l})j}^{mn} {\hat \theta}_{mn}^{{\hat l}j} = 0 \label{eq3.32c} 
\end{equation}
\begin{equation}
{\hat \theta}_{mn}^{{\hat \jmath}j} = 0 \label{eq3.32d}
\end{equation}
where ${\hat \theta}_{mn}^{{\hat \jmath}j}$ is the extended 
$P$-gravitational stress tensor. The extended tracelessness conditions in 
(3.32c) follow from (3.32a), and are independent of the equations of 
motion in Eq. (3.32d). Not surprisingly, the $P$-gravitational equations 
of motion reduce in the conformal gauge (3.31) to the following $2K$ 
extended Polyakov constraints:
\renewcommand{\theequation}{\thesection.\arabic{subsection}\alph{equation}}
\setcounter{subsection}{33}
\setcounter{equation}{0}
\begin{equation}
{\hat \theta}_{{\hat \jmath}j} = \tfrac {1}{4\pi} \tfrac {G^{ab}}{f_j(\sigma)} \sum_{{\hat l}=0}^{f_j(\sigma)-1} {\hat J}_{{\hat l}aj} {\hat J}_{{\hat \jmath}-{\hat l},bj} = 0 \label{eq3.33a} 
\end{equation}
\begin{equation}
{\hat {\bar \theta}}_{{\hat \jmath}j} = \tfrac {1}{4\pi} \tfrac {G^{ab}}{f_j(\sigma)} \sum_{{\hat l}=0}^{f_j(\sigma)-1} {\hat {\bar J}}_{{\hat l}aj} {\hat {\bar J}}_{{\hat \jmath}-{\hat l},bj} = 0 \label{eq3.33b} 
\end{equation}
\begin{equation}
{\hat J}_{{\hat \jmath}aj} = \tfrac {f_j(\sigma)}{2} G_{ab} \partial_+ {\hat x}^{-{\hat \jmath},bj},\ {\hat {\bar J}}_{{\hat \jmath}aj} = -\tfrac {f_j(\sigma)}{2} G_{ab} \partial_- {\hat x}^{-{\hat \jmath},bj} \label{eq3.33c} 
\end{equation}
\begin{equation}
\partial_- {\hat J}_{{\hat \jmath}aj} = \partial_+ {\hat {\bar J}}_{{\hat \jmath}aj} = \partial_- {\hat \theta}_{{\hat \jmath}j} = \partial_+ {\hat {\bar \theta}}_{{\hat \jmath}j} = 0 \label{eq3.33d} 
\end{equation}
\begin{equation}
{\bar {\hat \jmath}} = 0,\dots,f_j(\sigma)-1,\,\, \sum_j f_j(\sigma)=K,\,\, a,b = 1,\dots,d. \label{eq3.33e}
\end{equation}
These constraints are independent of the Weyl degrees of freedom, and are 
indeed nothing but the coordinate-space form of the extended Virasoro 
constraints (3.8) of the Hamiltonian formulation. I remind that these $2K$
constraints include the two ordinary Polyakov constraints
\renewcommand{\theequation}{\thesection.\arabic{equation}}
\setcounter{equation}{33}
\begin{equation}
\label{eq3.34}
\sum_j {\hat \theta}_{0j} = \sum_j {\hat {\bar \theta}}_{0j} = 0
\end{equation}
which involve only the total left- and right-mover physical stress 
tensors of each orbifold sector.

\setcounter{subsection}{3}
\subsection{Extended Nambu Actions}
\label{sec3.4}

For each twisted sector $\sigma$ of the permutation orbifolds $U(1)^{Kd}/ 
H(\mbox{perm})$, I also mention the extended action of Nambu-type:
\renewcommand{\theequation}{\thesection.\arabic{subsection}\alph{equation}}
\setcounter{subsection}{35}
\setcounter{equation}{0}
\begin{equation}
{\hat {\hat S}}_{\sigma} \approx \int dt \int_0^{2\pi} d\xi\, \sum_j {\hat {\hat H}}^{(0)j} \label{eq3.35a} 
\end{equation}
\begin{equation}
{\hat {\hat H}}^{({\hat \jmath})j} \equiv \tfrac {1}{f_j(\sigma)} \sum_{J=0}^{f_j(\sigma)-1} e^{-2\pi i \frac {{\hat \jmath}J}{f_j(\sigma)}} \left(-\det \left( \sum_{{\hat l}=0}^{f_j(\sigma)-1} e^{2\pi i \frac {J{\hat l}}{f_j(\sigma)}} {\hat {\mathcal L}}_{mn}^{{\hat l}j}\right)\right)^{\frac {1}{2}} \label{eq3.35b} 
\end{equation}
\begin{equation}
{\hat {\mathcal L}}_{mn}^{{\hat \jmath}j} =
 \tfrac {1}{8\pi} \sum_{{\hat l}=0}^{f_j(\sigma)-1} \partial_m {\hat x}^{{\hat l}aj} f_j(\sigma) G_{ab} \partial_n {\hat x}^{{\hat \jmath}-{\hat l},bj}. \label{eq3.35c}
\end{equation}
These systems are constructed using the quantity ${\hat {\mathcal 
L}}_{mn}^{{\hat \jmath}j}$ in Eq. (3.35c) as the induced, extended 
world-sheet metric, which transforms like the extended Polyakov metric:
\renewcommand{\theequation}{\thesection.\arabic{subsection}\alph{equation}}
\setcounter{subsection}{36}
\setcounter{equation}{0}
\begin{equation}
{\hat {\mathcal L}}_{mn}^{{\hat \jmath}j}(\xi + 2\pi,t) = e^{2\pi i \frac {\hat \jmath}{f_j(\sigma)}} {\hat {\mathcal L}}_{mn}^{{\hat \jmath}j}(\xi,t) \label{eq3.36a} 
\end{equation}
\begin{equation}
\delta {\hat {\mathcal L}}_{mn}^{{\hat \jmath}j} = \sum_{{\hat l}=0}^{f_j(\sigma)-1} ({\hat \beta}^{p{\hat l}j} \partial_p {\hat {\mathcal L}}_{mn}^{{\hat \jmath}-{\hat l},j} + \partial_m {\hat \beta}^{p{\hat l}j} {\hat {\mathcal L}}_{pn}^{{\hat \jmath}-{\hat l},j} + \partial_n {\hat \beta}^{p{\hat l}j} {\hat {\mathcal L}}_{pm}^{{\hat \jmath}-{\hat l},j}). \label{eq3.36b} 
\end{equation}
The transformations (3.36) follow from the monodromies (3.20c) and extended 
diffeomorphisms (3.22a) of the extended coordinates $\{{\hat x}^{{\hat\jmath}aj}, 
a = 1,\dots,d \}$, and imply in turn that the quantity ${\hat {\hat 
H}}^{({\hat \jmath})j}$transforms (like ${\hat H}^{({\hat\jmath})j}$ in (3.27)) as a
set of extended scalar densities.

 In particular, one finds that 
\renewcommand{\theequation}{\thesection.\arabic{subsection}\alph{equation}}
\setcounter{subsection}{37}
\setcounter{equation}{0}
\begin{equation}
{\hat {\hat H}}^{(0)j}(\xi + 2\pi,t)\, = \,{\hat {\hat H}}^{(0)j}(\xi,t),\quad\quad \sum_j f_j(\sigma) = K \label{eq3.37a}
\end{equation}
\begin{equation} 
\delta {\hat {\hat H}}^{(0)j} = \,\partial_p \left( \sum_{{\hat j}=0}^{f_j(\sigma)-1} {\hat \beta}^{p{\hat \jmath}j} {\hat {\hat H}}^{(0)j}\right) \label{eq3.37b}
\end{equation}
\begin{equation} 
\delta{\hat {\hat S}}_{\sigma} = 0 \label{eq3.37c}
\end{equation}
so the extended actions of Nambu-type are properly invariant under 
monodromy and the general permutation-twisted diffeomorphisms.

In Minkowski target space $G = \bigl(\begin{smallmatrix} -1 & 0 \\ 0 & 1\!\!1 
\end{smallmatrix}\bigr)$, the $N_* = 2K$ degrees of freedom of the extended 
diffeomorphisms should allow us to follow Ref. [66] in choosing the following 
$K$ gauge conditions as an {\it extended light-cone gauge}
\renewcommand{\theequation}{\thesection.\arabic{equation}}
\setcounter{equation}{37}
\begin{equation}
\label{eq3.38}
{\hat x}^{{\hat \jmath},+,j} = 0,\quad\quad {\bar {\hat \jmath}} = 0,\dots,f_j(\sigma)-1,\quad \sum_j f_j(\sigma) = K
\end{equation}
and show that the $K$ longitudinal coordinates ${\hat x}^{{\hat \jmath},-,j} = {\hat x}^{j,-,j}({\hat x}^{\perp})$
are dependent on the remaining $K(d-2)$ transverse twisted coordinates
 $\{{\hat x}^{{\hat \jmath}\alpha j}$, $\alpha = 1,\dots,d-2\}$. 
 Correspondingly, the extended light-cone gauge for the Nambu-like action 
 (2.58) of the orientation-orbifold sectors should give a description of 
 the twisted open strings in terms of $2(d-2)$ transverse degrees of freedom.
 
Subsec. 4.1 includes some remarks on the corresponding operator theories in critical dimension $d 
 = 26$.

\setcounter{subsection}{4}
\subsection{A Complementary Derivation}
\label{sec3.5}

Following our discussion in Subsec. 2.9 we are led to consider, in each 
sector of each permutation orbifold, a corresponding change of variable to 
the $J$-basis fields ${\hat A}^J, {\hat A}_J$ with twisted boundary 
conditions [29,37]:
\renewcommand{\theequation}{\thesection.\arabic{subsection}\alph{equation}}
\setcounter{subsection}{39}
\setcounter{equation}{0}
\begin{eqnarray}
{\hat x}^{{\hat \jmath}aj} &= &\tfrac {1}{f_j(\sigma)} \sum_{J=0}^{f_j(\sigma)-1} e^{-2\pi i \frac {{\hat \jmath}J}{f_j(\sigma)}} {\hat x}^{Jaj} \label{eq3.39a} \\
{\hat h}_{mn}^{({\hat \jmath})j} &= &\tfrac {1}{f_j(\sigma)} \sum_{J=0}^{f_j(\sigma)-1} e^{-2\pi i \frac {{\hat \jmath}J}{f_j(\sigma)}} {\hat h}_{mn}^{Jj} \label{eq3.39b} \\
{\hat h}_{({\hat \jmath})j}^{mn} &= &\tfrac {1}{f_j(\sigma)} \sum_{J=0}^{f_j(\sigma)-1} e^{2\pi i \frac {{\hat \jmath}J}{f_j(\sigma)}} {\hat h}_{Jj}^{mn} \label{eq3.39c} \\
{\hat H}^{({\hat \jmath})j} &= &\tfrac {1}{f_j(\sigma)} \sum_{J=0}^{f_j(\sigma)-1} e^{-2\pi i \frac {{\hat \jmath}J}{f_j(\sigma)}} \sqrt{-\det({\hat h}_{mn}^{Jj})} \label{eq3.39d} \\
{\hat h}_{mp}^{Jj} {\hat h}_{Jj}^{pn} &= &\delta_m^n,\quad J = 0,\dots,f_j(\sigma)-1,\quad \sum_j f_j(\sigma) = K. \label{eq3.39e}
\end{eqnarray}
Here $J$ is Fourier-conjugate to $\hat\jmath$, and the new fields are 
periodic $J \to J \pm f_j(\sigma)$, but the monodromies of the new fields 
are {\it not diagonal}
\renewcommand{\theequation}{\thesection.\arabic{subsection}\alph{equation}}
\setcounter{subsection}{40}
\setcounter{equation}{0}
\begin{equation}
{\hat A}^{Jj} = \sum_{{\hat \jmath}=0}^{f_j(\sigma)-1} e^{2\pi i \frac {J{\hat \jmath}}{f_j(\sigma)}} {\hat A}^{({\hat \jmath})j},\quad  {\hat A}_{Jj} = \sum_{{\hat \jmath}=0}^{f_j(\sigma)-1} e^{-2\pi i \frac {J{\hat \jmath}}{f_j(\sigma)}} {\hat A}_{({\hat \jmath})j} \label{eq3.40a} 
\end{equation}
\begin{equation}
{\hat A}^{Jj}(\xi + 2\pi,t) = {\hat A}^{J+1,j}(\xi,t),\quad {\hat A}_{Jj}(\xi+2\pi,t) = {\hat A}_{J+1,j}(\xi,t). \label{eq3.40b}
\end{equation}
After some algebra, one finds that the general extended action (3.28) of 
Polyakov-type takes the following simple form in the $J$-basis
\renewcommand{\theequation}{\thesection.\arabic{subsection}\alph{equation}}
\setcounter{subsection}{41}
\setcounter{equation}{0}
\begin{equation}
{\hat S}_{\sigma} = \int dt \int_0^{2\pi} d\xi\,\, \tfrac {1}{8\pi} \sum_j \sum_{{\hat \jmath},{\hat k},{\hat l}=0}^{f_j(\sigma)-1} {\hat H}^{({\hat \jmath})j} {\hat h}_{({\hat \jmath}+{\hat k}+{\hat l})j}^{mn} \partial_m {\hat x}^{{\hat k}aj} f_j(\sigma) G_{ab}\partial_n {\hat x}^{{\hat l}bj} \label{eq3.41a}
\end{equation}
\begin{equation}
\hspace{.3in}= \int dt \int_0^{2\pi} d\xi\,\, \tfrac {1}{8\pi} \sum_j \sum_{J=0}^{f_j(\sigma)-1} \sqrt{-\det({\hat h}^{Jj})}\,\, {\hat h}_{Jj}^{mn} \partial_m {\hat x}^{Jaj} G_{ab} \partial_n {\hat x}^{Jbj} \label{eq3.41b} 
\end{equation}
for each sector $\sigma$ of each permutation orbifold. Because of the sum on $J$, 
the action density in Eq. (3.41b) is still manifestly monodromy-invariant 
under the non-diagonal monodromies of the $J$-basis fields. Moreover -- 
except for global coupling via the non-diagonal monodromies -- the form (3.41b) is a sum of $K$ 
ordinary untwisted Polyakov actions [64] for the original closed-string CFT $U(1)^{Kd}$, 
written now in the cycle basis of each representative element $h_{\sigma} \in H(\mbox{perm})$.

This result allows us to understand the general extended action (3.41a) 
of Polyakov-type as nothing but the form obtained by monodromy- 
decomposition of the $J$-basis fields $\hat A^{Jj}, \hat A_{Jj}$, an 
essentially standard application of the principle of local isomorphisms 
[27,29,31,32,35,37]. Using only Eq. (3.39a) for the twisted coordinates $\{{\hat x}^{{\hat \jmath}aj}\}$, the extended 
actions (3.35) of Nambu-type can similarly be understood as the 
monodromy-resolved form of a sum of $K$ ordinary untwisted Nambu actions [65] for the 
mixed-monodromy coordinates $\{{\hat x}^{Jaj}\}$. 

Another useful form of the $J$-basis action (3.41b) is
\renewcommand{\theequation}{\thesection.\arabic{equation}}
\setcounter{equation}{41}
\begin{equation}
\label{eq3.42}
{\hat S}_{\sigma} = \sum_j \sum_{J=0}^{f_j(\sigma)-1} \int dt_{Jj} \int_0^{2\pi} d\xi_{Jj}\, \tfrac {1}{8\pi} \sqrt{-\det({\hat h}_{mn}^{Jj})}\,\, {\hat h}_{Jj}^{mn} \partial_m {\hat x}^{Jaj} G_{ab} \partial_n {\hat x}^{Jbj}
\end{equation}
where I have relabelled the dummy variables 
$\{t,\xi \rightarrow t_{Jj},\xi_{Jj}\}$ separately in each of the $K$ terms. 
Then we see that the action of sector $\sigma$ is locally 
invariant under the product of $K$ diffeomorphism groups
\renewcommand{\theequation}{\thesection.\arabic{subsection}\alph{equation}}
\setcounter{subsection}{43}
\setcounter{equation}{0}
\begin{equation}
\xi_{Jj}^{m\prime} = \xi_{Jj}^{m\prime} (\{\xi_{Jj}^p\}),\quad m = 0,1,\,  
J = 0,\dots,f_j(\sigma)-1,\ \sum_j f_j(\sigma)=K \label{3.43a}
\end{equation}
\begin{equation}
{\hat x}^{Jaj\prime}(\{\xi_{Jj}^{p\prime}\}) = {\hat x}^{Jaj}(\{\xi_{Jj}^p\}),\quad a = 0,\dots,d-1 \label{eq3.43b} 
\end{equation}
\begin{equation}
{\hat h}_{mn}^{Jj\prime}(\{\xi_{Jj}^{p\prime}\}) = \tfrac {\partial\xi_{Jj}^p}{\partial\xi_{Jj}^{m\prime}} \tfrac {\partial\xi_{Jj}^q}{\partial\xi_{Jj}^{n\prime}} {\hat h}_{pq}^{Jj}(\{\xi_{Jj}^p\}) \label{eq3.43c}
\end{equation}
which are {\it globally intertwined} by the non-diagonal 
monodromies of the $J$-basis.

In principle, all the invariances of the $J$-basis can be mapped back 
into the twisted $\hat\jmath$-basis via the monodromy-decompositions 
(3.39). I illustrate this here with the simple case of the extended Weyl 
invariance, again leaving the finite form of the twisted diffeomorphism 
groups for future work.

The Weyl invariance of ${\hat S}_{\sigma}$ in the $J$-basis is
\renewcommand{\theequation}{\thesection.\arabic{equation}}
\setcounter{equation}{43}
\begin{equation}
\label{eq3.44}
{\hat h}_{mn}^{Jj} \to e^{-{\hat \sigma}_{Jj}} {\hat h}_{mn}^{Jj},\quad {\hat \sigma}_{Jj}(\xi+2\pi,t) = {\hat \sigma}_{J+1,j}(\xi,t).
\end{equation}
Then, using the monodromy decomposition (3.39) of the extended metric and
\renewcommand{\theequation}{\thesection.\arabic{equation}}
\setcounter{equation}{44}
\begin{equation}
\label{eq3.45}
{\hat \sigma}_{({\hat \jmath})j} \equiv \tfrac {1}{f_j(\sigma)} \sum_{J=0}^{f_j(\sigma)-1} e^{2\pi i \frac {{\hat \jmath}J}{f_j(\sigma)}} {\hat \sigma}_{Jj}
\end{equation}
one finds the extended, twisted Weyl invariance in the monodromy-resolved 
$\hat \jmath$-basis:
\renewcommand{\theequation}{\thesection.\arabic{subsection}\alph{equation}}
\setcounter{subsection}{46}
\setcounter{equation}{0}
\begin{equation}
{\hat h}_{mn}^{({\hat \jmath})j} \to \tfrac {1}{f_j(\sigma)} \sum_{J,{\hat k}=0}^{f_j(\sigma)-1} \exp\left( 2\pi i \tfrac {J({\hat k}-{\hat \jmath})}{f_j(\sigma)} - \sum_{{\hat l}=0}^{f_j(\sigma)-1} e^{-2\pi i \frac {J{\hat l}}{f_j(\sigma)}} {\hat \sigma}_{({\hat l})j}\right) {\hat h}_{mn}^{({\hat k})j} \label{eq3.46a} 
\end{equation}
\begin{equation}
{\hat \sigma}_{({\hat \jmath})j}(\xi+2\pi,t) = e^{-2\pi i\frac {\hat \jmath}{f_j(\sigma)}} {\hat \sigma}_{({\hat \jmath})j} (\xi,t) \label{eq3.46b}
\end{equation}
\begin{equation}
{\tilde h}_{({\hat \jmath})j}^{mn} \to {\tilde h}_{({\hat \jmath})j}^{mn},\quad {\hat S}_{\sigma} \to {\hat S}_{\sigma}. \label{eq3.45c}
\end{equation}
Similarly, the standard conformal-gauge parametrization of the metric in 
the $J$-basis
\renewcommand{\theequation}{\thesection.\arabic{subsection}\alph{equation}}
\setcounter{subsection}{47}
\setcounter{equation}{0}
\begin{equation}
{\hat h}_{mn}^{Jj} = \eta_{mn} e^{-{\hat \phi}_{Jj}},\quad {\hat \phi}_{Jj}(\xi+2\pi,t) = {\hat \phi}_{J+1,j}(\xi,t) \label{eq3.47a} 
\end{equation}
\begin{equation}
{\hat \phi}_{({\hat \jmath})j} \equiv \tfrac {1}{f_j(\sigma)} \sum_{J=0}^{f_j(\sigma)-1} e^{2\pi i \frac {{\hat \jmath}J}{f_j(\sigma)}} {\hat \phi}_{Jj} \label{eq3.47b}
\end{equation}
gives the promised conformal-gauge result (3.31) in terms of the $K$ extended 
Weyl fields $\{{\hat \phi}_{({\hat \jmath})j}\}$ with diagonal monodromy.

As a simple example consider the case of $\mathbb{Z}_2$-permutation 
gravity, for which the monodromy-resolved results above read:
\renewcommand{\theequation}{\thesection.\arabic{subsection}\alph{equation}}
\setcounter{subsection}{48}
\setcounter{equation}{0}
\begin{equation}
{\hat h}_{mn}^{(0)} \to e^{-{\hat \sigma}_{(0)}} ({\hat h}_{mn}^{(0)}\cos({\hat \sigma}_{(1)}) - i {\hat h}_{mn}^{(1)}\sin({\hat \sigma}_{(1)})) \label{eq3.48a} 
\end{equation}
\begin{equation}
{\hat h}_{mn}^{(1)} \to e^{-{\hat \sigma}_{(0)}} ({\hat h}_{mn}^{(1)}\cos({\hat \sigma}_{(1)}) - i {\hat h}_{mn}^{(0)}\sin({\hat \sigma}_{(1)})) \label{eq3.48b} 
\end{equation}
\begin{equation}
{\hat h}_{mn}^{(0)} = \eta_{mn} e^{-{\hat \phi}_{(0)}} \cos({\hat \phi}_{(1)}),\quad {\hat h}_{mn}^{(1)} = -i \eta_{mn} e^{-{\hat \phi}_{(0)}}\sin({\hat \phi}_{(1)}) \label{eq3.48c} 
\end{equation}
\begin{equation}
{\hat \sigma}_{(u)}(\xi + 2\pi,t) = (-1)^u{\hat \sigma}_{(u)}(\xi,t),\,\,\,\, {\hat \phi}_{(u)}(\xi + 2\pi,t) = (-1)^u {\hat \phi}_{(u)}(\xi,t)  \label{eq3.48d} 
\end{equation}
\begin{equation}
{\hat h}_{mn}^{(u)}(\xi+2\pi,t) = (-1)^u {\hat h}_{mn}^{(u)}(\xi,t). \label{eq3.48e}
\end{equation}
Here I have suppressed the single cycle index $j = 0$, and relabelled ${\bar {\hat \jmath}} = {\bar u} = 0,1$.
These are the same extended Weyl transformations (2.62b) and conformal-gauge 
extended metric (2.51b) found for $\mathbb{Z}_2$-permutation gravity in the open-string 
orientation-orbifold sectors, except that those results were left in 
terms of the fields with non-diagonal monodromy ($0 \!\leftrightarrow\! 1$)
\renewcommand{\theequation}{\thesection.\arabic{equation}}
\setcounter{equation}{48}
\begin{equation}
\label{eq3.49}
{\hat \sigma}_0 = {\hat \sigma}_{(0)} + {\hat \sigma}_{(1)},\ {\hat \sigma}_1 = {\hat \sigma}_{(0)} - {\hat \sigma}_{(1)},\ {\hat \phi}_0 = {\hat \phi}_{(0)} + {\hat \phi}_{(1)},\ {\hat \phi}_1 = {\hat \phi}_{(0)} - {\hat \phi}_{(1)}
\end{equation}
which would be inappropriate for the final form of the permutation orbifolds.

\setcounter{section}{3}
\section{Discussion}
\label{sec4}

\setcounter{subsection}{0}
\subsection{The Conjecture: \\
 \hspace*{.2in} Physical Strings at Higher Central Charge}
\label{sec4.1}

Based on their extended (twisted) Virasoro algebras [27,55,35], I have found 
extended world-sheet action formulations, of both Polyakov- and 
Nambu-type, in the twisted sectors of the orbifolds of permutation-type.
This includes in particular the twisted open-string sectors at $\hat 
c=2d$ of the orientation orbifolds $U(1)^d/ ({\mathbb Z}_2(w.s.) \times 
H)$ and the twisted closed-string sectors at $\hat c=Kd$ of the permutation orbifolds $U(1)^{Kd}/ H(\mbox{perm})$,
where in both cases $U(1)^d$ is the d-dimensional free-bosonic closed-string  CFT.
The extended actions of Polyakov-type exhibit a class of new extended (twisted)
 world-sheet gravities called the {\it permutation gravities}, which are classified by the conjugacy classes of
the permutation groups. Both the Polyakov- and the Nambu-type actions are invariant under 
the extended, twisted diffeomorphism groups generated by the extended 
Virasoro algebras.

In the covariant formulation of the corresponding Minkowski-space quantum 
theories, each of these twisted sectors carries an increased number of 
{\it negative-norm states} (ghosts) corresponding to the higher central 
charge. The negative norms are associated as usual with the number of 
 twisted time-like currents of each sector:
\renewcommand{\theequation}{\thesection.\arabic{equation}}
\setcounter{equation}{0}
\begin{equation}
\label{eq4.1}
N_* = \left\{ \begin{array}{rl}
2 &\mbox{for open-string orient.orb.sectors} \\
2K &\mbox{for closed-string perm.orb.sectors.}
\end{array} \right.
\end{equation}
For example, the $2K$ twisted time-like currents of the permutation 
orbifolds have the form
\renewcommand{\theequation}{\thesection.\arabic{subsection}\alph{equation}}
\setcounter{subsection}{2}
\setcounter{equation}{0}
\begin{equation}
\partial_{\pm}{\hat x}^{{\hat \jmath}0j},\quad {\bar {\hat j}} = 0,1,\dots,f_j(\sigma)-1,\quad \sum_j f_j(\sigma) = K \label{eq4.2a} 
\end{equation}
\begin{equation}
G_{ab} = \begin{pmatrix} -1 & 0 \\ 0 & 1\!\!1 \end{pmatrix},\quad a = 0,1,\dots,d-1 \label{eq4.2b}
\end{equation}
and (although it is masked by the unitary transformation (2.23) of the 
Minkowski metric) the ghost-doubling in (4.1) for the orientation 
orbifolds follows in the same way from the doubling $\bar u=0,1$ of the 
extended coordinates. In both cases, the counting (4.1) of time-like 
currents holds as well for the untwisted sector of the orbifolds -- where the
doubling for the orientation orbifolds counts both the left- and 
right-mover currents of the untwisted closed-string CFT.

But we have seen the number $N_{*}$ earlier in our discussion. It is also the number of 
degrees of freedom in the extended diffeomorphisms of each twisted 
sector, or equivalently the number of extended Virasoro (Polyakov) constraints
(see Eqs. (2.56) and (3.9)) in each sector. This allows us to conjecture that the orbifolds
of permutation-type define operator-string theories which are free of 
negative-norm states at higher central charge. In particular, we may 
expect that the extended Virasoro constraints of the classical theories will
translate into {\it extended physical-state conditions} associated to extended Ward identities for the
twisted tree amplitudes and 
loops of the corresponding orbifold-string theories. 

More precisely, consideration of orbifold-string loops and/or 
BRST operators for the twisted sectors should fix the critical dimension 
of the ghost-free theories at $d=26$, so that the critical orbifold-string theories
\renewcommand{\theequation}{\thesection.\arabic{subsection}\alph{equation}}
\setcounter{subsection}{3}
\setcounter{equation}{0}
\begin{equation}
\frac {U(1)^{26}}{{\mathbb Z}_2(w.s.) \times H},\quad {\hat c} = 52 \mbox{ for open-string sectors} \label{eq4.3a} 
\end{equation}
\begin{equation}
\frac {U(1)^{26K}}{H(\mbox{perm})},\quad {\hat c} = 26K \mbox{ for all sectors} \label{eq4.3b}
\end{equation}
are orbifolds of decoupled copies of the critical free-bosonic string. 
This picture makes sense because orbifolding should not create 
negative-norm states when the original symmetric string theory was completely physical.

In the succeeding papers of this series, I will augment this intuition by 
constructing the twisted BRST systems of $\hat c=52$ matter, as well as new
extended Ward identities, ghost-free tree amplitudes and 
modular-invariant loops for the permutation orbifold-strings. I will also be 
able to shed some light on the somewhat mysterious relation between 
orientation orbifolds and conventional orientifolds.

The extended actions of Nambu-type offer another approach to this 
conjecture. In this case one can choose an extended light-cone gauge (see 
Subsec. 3.4) in each twisted sector, which should then show effective 
central charges
\renewcommand{\theequation}{\thesection.\arabic{equation}}
\setcounter{equation}{3}
\begin{equation}
\label{eq4.4}
{\hat c}_{eff} = \left\{ \begin{array}{rl}
52 - 4 = (26-2) \cdot 2 = 48 \,\,\,&\,\,\,\mbox{(orientation orbs)} \\
(26-2)K = 24K &\,\,\,\mbox{(permutation orbs)}
\end{array} \right.
\end{equation}
for the physical (transverse) degrees of freedom in the critical 
orbifolds. Following Ref. [66] we know that consistency of this 
quantization will involve space-time interpretation of the extended 
coordinates $\hat x$ (52 or 26+26 ?), to be determined by closure of the 
appropriate space-time group generators in each twisted sector.

\setcounter{subsection}{1}
\subsection{Other Orbifolds of Permutation-type}
\label{sec4.2}

Our classical discussion of the orbifolds of permutation-type is by no 
means complete. Beyond the issues left unfinished here for the 
orientation and permutation orbifolds, many other orbifolds of permutation-type
are known, whose extended formulation can be studied with the 
techniques developed here:

\vspace{.1in}\noindent $\bullet$ The {\it generalized} free-bosonic permutation orbifolds [45,46] 
\renewcommand{\theequation}{\thesection.\arabic{equation}}
\setcounter{equation}{4}
\begin{equation}
\label{eq4.5}
\frac {U(1)^{26K}}{H_+},\,\,\, H_+ = H(\mbox{perm}) \times H'
\end{equation}
at critical central charge $\hat c=26K$. Here the group $H_+$ can involve 
extra automorphisms $H'$ which act uniformly on each closed-string copy 
$U(1)^{26}$. Following the reasoning above, I present only the initial 
data and the final form of the extended action for each twisted sector of the
generalized permutation orbifolds:
\renewcommand{\theequation}{\thesection.\arabic{subsection}\alph{equation}}
\setcounter{subsection}{6}
\setcounter{equation}{0}
\begin{equation}
\left. J_{aI}\right. \!' = P_I{}^J \omega_a{}^bJ_{bJ},\,\, 
\left.\bar{J}_{aI}\right. \!' = P_I{}^J \omega_a{}^b {\bar J}_{bJ} \label{eq4.6a}
\end{equation}
\begin{equation}
P \in H(\mbox{perm}),\quad \omega \in H' \label{eq4.6b} 
\end{equation}
\begin{equation}
\omega_a{}^b(U^{\dag})_b{}^{n(r)\mu} = (U^{\dag})_a{}^{n(r)\mu} e^{-2\pi i \frac {n(r)}{\rho(\sigma)}} \label{eq4.6c} 
\end{equation}
\begin{equation}
{\mathcal G}_{n(r)\mu;n(s)\nu} \equiv \chi_{n(r)\mu} \chi_{n(s)\nu} U_{n(r)\mu}{}^a U_{n(s)\nu}{}^b G_{ab} \label{eq4.6d} 
\end{equation}
\begin{equation}
= \delta_{n(r)+n(s),0 \!\!\!\!\mod \rho(\sigma)} {\mathcal G}_{n(r)\mu;-n(r)\nu} \label{eq4.6e} 
\end{equation}
\begin{equation}
\hspace{-3.4in}{\hat S}_{\sigma} = \tfrac {1}{8\pi} \int dt \int_0^{2\pi} d\xi\,\, \times\label{eq4.6f} 
\end{equation}
\begin{equation}
\hspace{.2in}\times\sum_j f_j(\sigma) \sum_{{\hat \jmath},{\hat k},{\hat l}=0}^{f_j(\sigma)-1} {\hat H}^{({\hat \jmath})j} {\hat h}_{({\hat \jmath}+{\hat k}+{\hat l})j}^{mn} \partial_m {\hat x}^{{\hat k}n(r)\mu j} {\mathcal G}_{n(r)\mu;n(s)\nu} \partial_n {\hat x}^{{\hat l}n(s)\nu j} \nonumber
\end{equation}
\begin{equation}
{\hat x}^{{\hat \jmath}n(r)\mu j}(\xi + 2\pi,t) = e^{-2\pi i \left( \frac {\hat \jmath}{f_j(\sigma)} + \frac {n(r)}{\rho(\sigma)}\right)} {\hat x}^{{\hat \jmath}n(r)\mu j} (\xi,t) \label{eq4.6g}
\end{equation}
\begin{equation}
{\bar {\hat \jmath}} = 0,1,\dots,f_j(\sigma)-1,\quad \sum_j f_j(\sigma) = 
K,\quad  {\bar n}(r) \in (0,1,\dots,\rho(\sigma)-1). \label{eq4.6h} 
\end{equation}
In these cases, the twisted metric  $\mathcal G$ in Eq. (4.6d) is constructed from the 
solution of the H-eigenvalue problem (4.6c) of each extra automorphism $\omega \in H'$. 
Note in particular that, as seen in the extended actions (4.6f),
the generalized permutation 
orbifolds involve the {\it same} twisted permutation gravities studied 
above. 

We have seen such a universality before in the case of the $\mathbb{Z}_2$-permutation 
gravity of all the orientation orbifolds. Indeed the permutation gravities couple 
universally to the extended stress tensors,
which are associated to the same orbifold Virasoro algebras (3.2) in each twisted 
sector of any orbifold of permutation-type.
The orbifold Virasoro algebras are themselves universal because 
they arise [35,43] simply
  by twisting the action of $H(\mbox{perm})$ on the ordinary decoupled Virasoro copies 
 in the untwisted sector of the orbifold -- independent of the action of any 
 particular $H'$ on a given copy. 
 For the generalized permutation orbifolds in Eq. (4.6), the explicit form of 
 their ``doubly-twisted'' current algebras (with modeing 
 $J_{\hat jn(r)\mu j}(m+\tfrac{\hat j}{f_{j}(\sigma)}+\tfrac{n(r)}{\rho(\sigma)}$) and (conformal gauge)
extended stress tensors 
can be obtained by the substitution $\{a\rightarrow n(r)\mu, 
G\rightarrow \mathcal{G}\}$ in Eqs. (3.14a,d,e).

\vspace{.1in}\noindent $\bullet$ The free-bosonic {\it open-string 
permutation orbifolds} [45]
\renewcommand{\theequation}{\thesection.\arabic{equation}}
\setcounter{equation}{6}
\begin{equation}
\label{eq4.7}
\frac {U(1)_{\mbox{open}}^{26K}}{H(\mbox{perm})}
\end{equation}
and their T-dualizations [46] at critical central charge $\hat c=26K$, 
which will also be governed by the same permutation gravities. Starting 
from the left-mover data of the generalized (closed-string) permutation 
orbifolds above, the generalized open-string permutation orbifolds
\renewcommand{\theequation}{\thesection.\arabic{equation}}
\setcounter{equation}{7}
\begin{equation}
\label{eq4.8}
\frac {U(1)_{\mbox{open}}^{26K}}{H(\mbox{perm}) \times H'}
\end{equation}
and their T-duals at ${\hat c} = 26K$  can also be worked out as special cases 
of Ref. [46]. (The sectors of the generalized open-string 
$\mathbb{Z}_2$-permutation orbifolds are T-dual to the open-string 
sectors of the orientation orbifolds.)

\vspace{.1in}\noindent $\bullet$ The {\it superstring orbifolds of permutation-type} at 
critical central charge $\hat c=10K$ (and  $({\hat c},{\hat {\bar c}}) = 
(26K,10K)$ for heterotic type). The first goal here will be the explicit form of the
world-sheet {\it permutation supergravities}, associated to the extended, 
twisted superconformal algebras [22,27] of these orbifolds.

\vspace{.1in}\noindent $\bullet$ The {\it partial orbifoldizations} of permutation-type, 
for example
\renewcommand{\theequation}{\thesection.\arabic{equation}}
\setcounter{equation}{8}
\begin{equation}
\label{eq4.9}
\frac {U(1)^{26} \times U(1)^{26}}{{\mathbb Z}_2(D)},\quad 0 \le D \le 25,\quad {\hat c} = 52
\end{equation}
where $\mathbb{Z}_{2}(D)$ exchanges only two subsets $\{D\}$ of $D$ bosons each. 
The single twisted sector of this orbifold is described by a {\it hybrid} action
\renewcommand{\theequation}{\thesection.\arabic{equation}}
\setcounter{equation}{9}
\begin{equation}
\label{eq4.10}
{\hat S} = S_1(26-D) + S_2(26-D) + {\hat S}(2D)
\end{equation}
where $S_{1,2}$ are ordinary Polyakov actions for $26-D$ untwisted bosons 
and $\hat S(2D)$ is the extended action (3.28) with $\mathbb{Z}_2$-twisted 
permutation gravity coupled to $2D$ twisted bosons
\renewcommand{\theequation}{\thesection.\arabic{equation}}
\setcounter{equation}{10}
\begin{equation}
\label{eq4.11}
\{{\hat x}^{{\hat \jmath}a0},\\, \forall\ a \in \{D\},\, {\bar {\hat \jmath}} = 0,1\}.
\end{equation}
Extensions to higher genus, as well as non-trivial $B$ fields and twisted
$B$ fields [37] can also be studied. 

Our discussion above suggests that  all the critical orbifold CFT's 
of permutation-type can describe twisted physical string systems at higher 
central charge.

\section*{Acknowledgements}

For helpful information, discussions and encouragement,I thank L. 
Alvarez-Gaum$\acute{e}$, K. Bardakci, I. Brunner,
J. de Boer, D. Fairlie, O. Ganor, E. Gimon, C. Helfgott, E. Kiritsis, R. 
Littlejohn, S. Mandelstam, J. McGreevy, N. Obers, A. Petkou, E. 
Rabinovici, V. Schomerus, K. Schoutens, C. Schweigert and E. Witten. This work was 
supported in part by the Director, Office of Energy Research, Office of 
High Energy and Nuclear Physics, Division of High Energy Physics of the U.S.
Department of Energy under Contract DE-AC02-O5CH11231 and in part by the National
Science Foundation under grant PHY00-98840.


\begin{thebibliography}{99}


\bibitem{KM} 
V.~Kac, ``Simple graded {Lie} algebras of finite growth,'' {\em Funct. Anal. 
  Appl.} {\bf 1} (1967) 328; R.~V. Moody, ``Lie algebras associated with generalized {Cartan} matrices,'' 
  {\em Bull. Am. Math. Soc.} {\bf 73} (1967) 217--221. 
 
\bibitem{BH} 
K.~Bardakci and M.~B. Halpern, ``New dual quark models,'' {\em Phys. Rev.} {\bf 
  D3} (1971) 
2493. 

\bibitem{2faces1} 
M.~B. Halpern, ``The two faces of a dual pion-quark model,'' {\em Phys. Rev.} 
  {\bf D4} (1971) 2398.
  
\bibitem{Dash} 
R.~Dashen and Y.~Frishman, ``Four-fermion interactions and scale invariance,'' 
{\em Phys. Rev.} {\bf D11} (1975) 2781. 

\bibitem{Su(n)_{1}}
M.~B. Halpern, ``Quantum `solitons' which are SU(N) fermions,'' {\em Phys. 
Rev.} {\bf D12} (1975) 1684; M.~B. Halpern, ``Equivalent-boson method and free currents in 
two-dimensional gauge theories,'' {\em Phys. Rev.} {\bf D13} (1976) 
337; T.~Banks, D.~Horn and H.~Neuberger, ``Bosonization of the SU(N) 
Thirring Models,'' {\em Nucl. Phys.} {\bf 
B108} (1976) 119; I.~B. Frenkel and V.~G. Kac, ``Basic representations of affine Lie 
algebras and dual resonance models,'' {\em Inv. Math.} {\bf 62} (1980) 23.

\bibitem{NovWitt}
S.~P.Novikov, ``The Hamiltonian formalism and a many-valued analogue 
of Morse theory,'' {\em Usp.Mat.Nauk} {\bf 37} (1982) 3;
E.~Witten, `` Nonabelian bosonization in two dimensions,'' {\em Comm. Math
Phys.} {\bf 92} (1984) 455. 
   
\bibitem{KZ} 
V.~G. Knizhnik and A.~B. Zamolodchikov, ``Current algebra and Wess-Zumino model in
   two dimensions,'' {\em Nucl. Phys.} {\bf B247} (1984) 83. 
 

\bibitem{GKO}
P.~Goddard, A.~Kent and D.~Olive, ``Virasoro algebras and coset-space 
models,'' {\em Phys. Lett.} {\bf B152} (1985) 88.

\bibitem{Cosact}
K.~Barkakci, E.~Rabinovici and B.~S$\ddot{a}$ring, ``String models 
with $c<1$ components,'' {\em Nucl. Phys.} 
{\bf B299} (1988) 151; K.~Gawedzki and A.~Kupiainen, ``G/H conformal 
field theory from  gauged WZW model,'' {\em Phys. Lett.} {\ B215} (1988)
 119; D.~Karabali, Q.-H. Park, H.~J. Schnitzer and Z.~Yang, ``A GKO 
 construction based on a path-integral formulation of gauged 
 Wess-Zumino-Witten actions,'' {\em Phys. 
Lett.} {\bf B216} (1989) 307; D.~Karabali and H.~J. Schnitzer, ``BRST 
quantization of a gauged WZW action and coset conformal field 
theories,''{\em Nucl. Phys.}
 {\bf B329} (1990) 649.

\bibitem{VME}
M.~B. Halpern and E.~Kiritsis, ``General Virasoro Construction on 
affine g,''  {\em Mod. Phys. Lett.}  {\bf A4} (1989) 1373; 
Erratum, ibid. {\bf A4} (1989) 1797. A.~Y. Morozov, A.~M. Perelemov, A.~A. Roslyi, M.~A. Shifman and A.~V. 
Turbiner, ``Quasiexactly solvable quantal problems: One-dimensional
analogue of rational conformal field theories,'' Int. J.Mod.Phys. 
A5 (1990) 803.

\bibitem{ICFT} 
M.~B. Halpern, E.~Kiritsis, N.~A. Obers and K.~Clubok, ``Irrational 
conformal field theory,'' {\em Phys. Rep.} {\bf 265} 
(1996) 1, hep-th/9501144. 

\bibitem{2faces2}
M.~B. Halpern and C.~B. Thorn, ``The two faces of a dual pion-quark model 
II. Fermions and other things,'' {\em Phys. Rev.} {\bf D4} (1971) 3084.

\bibitem{Corr}
E.~Corrigan and D.~B. Fairlie, ``Off-shell states in dual resonance 
theory,''  {\em Nucl. Phys.}{\bf B91} (1975) 527.

\bibitem{Warren}
W.~Siegel, ``Strings with dimension-dependent intercept,'' {\em Nucl. Phys.} {\bf B109} (1976) 244.

\bibitem{Lep}
J.~Lepowsky and R.~L. Wilson, ``Construction of the affine Lie 
algebra $A_{1}^{(1)}$,'' {\em Comm. Math. Phys.} {\bf 62} (1978) 43.

\bibitem{DixHarv}
L.~Dixon, J.~Harvey, C.~Vafa and E.~Witten, ``Strings on orbifolds,'' 
{\em Nucl. Phys.} {\bf B261} (1985) 678; ``Strings on orbifolds II,'' 
{\em Nucl. Phys.} {\bf B274} (1986) 285.

\bibitem{DixFried}
L.~Dixon, D.~Friedan, E.~Martinec and S.~Shenker, `` The conformal 
field theory of orbifolds,'' {\em Nucl. Phys.} {\bf B282} (1987) 13.

\bibitem{Ham}
S.~Hamidi and C.~Vafa, ``Interactions on orbifolds,'' {\em Nucl. 
Phys.} {\bf B279} (1987) 465.

\bibitem{Defs}
J.~K. Freericks and M.~B. Halpern, ``Conformal deformation by the currents 
of affine g,'' {\em Ann. Phys.} {\bf 188} (1988) 258.

\bibitem{DV3}
R.~Dijkgraff, C.~Vafa, E.~Verlinde and H.~Verlinde, `` The operator 
algebra of orbifold models,'' {\em Comm. Math. Phys.} {\bf 123} (1989) 485.

\bibitem{Klemm}
A.~Klemm and M.~G. Schmidt, ``Orbifolds by cyclic permutations of 
tensor-product conformal field theories, '' {\em Phys. Lett.} {\bf 
B245} (1990) 53.

\bibitem{Fuchs}
J.~Fuchs, A.~Klemm and M.~G. Schmidt, `` Orbifolds by cyclic 
permutations in Gepner-type superstrings and in the corresponding
Calabi-Yao manifolds,'' {\em Ann. Phys.} {\bf 214} (1992) 221.

\bibitem{Ven}
G.~Veneziano, ``Construction of a crossing-symmetric Regge-behaved 
amplitude for linearly rising trajectories,'' {\em Nuovo Cimento} {\bf 57A} (1968).

\bibitem{Closed}
M.~A. Virasoro, ``Alternative constructions of crossing-symmetric 
amplitudes with Regge behavior,'' {\em Phys. Rev.} {\bf 177} (1969) 
2309; J.~Shapiro, ``Electrostatic analogue for the Virasoro model,'' {\em Phys. 
Lett.} {\bf B33} (1970) 361.

\bibitem{Mand}
S.~Mandelstam, ``Dual resonance models,'' {\em Phys. Rep.} {\bf 13} (1974) 259.

\bibitem{GSW}
M.~B. Green, J.~H. Schwarz and E.~Witten, ``Superstring theory,''  
Cambridge University Press, 1987.

\bibitem{Christ}
L.~Borisov, M.~B. Halpern, and C.~Schweigert, ``Systematic approach to cyclic 
  orbifolds,'' {\em Int. J. Mod. Phys.} {\bf A13} (1998) 125, hep-th/9701061. 
 
\bibitem{OVME} 
J.~Evslin, M.~B. Halpern, and J.~E. Wang, ``General {Virasoro} construction on 
  orbifold affine algebra,'' {\em Int. J. Mod. Phys.} {\bf A14} (1999) 
  4985, hep-th/9904105. 
 
\bibitem{Dual} 
J.~de~Boer, J.~Evslin, M.~B. Halpern, and J.~E. Wang, ``New duality 
  transformations in orbifold theory,'' {\em Int. J. Mod. Phys.} {\bf A15} 
  (2000) 1297, hep-th/9908187. 

 \bibitem{Coset} 
J.~Evslin, M.~B. Halpern, and J.~E. Wang, ``Cyclic coset orbifolds,'' {\em Int. 
  J. Mod. Phys.} {\bf A15} (2000) 3829,  hep-th/9912084. 
 
\bibitem{More} 
M.~B. Halpern and J.~E. Wang, ``More about all current-algebraic orbifolds,'' 
  {\em Int. J. Mod. Phys.} {\bf A16} (2001) 97, hep-th/0005187. 
 
\bibitem{Big} 
J.~de~Boer, M.~B. Halpern, and N.~A. Obers, ``The operator algebra and twisted 
  {KZ} equations of {WZW} orbifolds,'' {\em J. High Energy Phys.} {\bf 10} (2001) 011, 
 hep-th/0105305. 
 
\bibitem{Big'} 
M.~B. Halpern and N.~A. Obers, ``Two large examples in orbifold theory: 
Abelian orbifolds and 
  the charge conjugation orbifold on $su(n)$,'' {\em Int. J. Mod. Phys.} {\bf A17} (2002) 3897, hep-th/0203056.

\bibitem{Fab} 
M.~B. Halpern and F.~Wagner, ``The general coset orbifold action,'' {\em Int. J. Mod. Phys.}
  {\bf A18} (2003) 19, hep-th/0205143. 

\bibitem{Perm}
M.~B. Halpern and C.~Helfgott, ``Extended operator algebra and reducibility in the WZW permutation orbifolds,''
    {\em Int. J. Mod. Phys.} {\bf A18} (2003) 1773,  hep-th/0208087. 
    
\bibitem{So2n}
O.~Ganor, M.~B. Halpern, C.~Helfgott and N.~A. Obers, ``The outer-automorphic 
WZW orbifolds on $so (2n)$, including
    five triality orbifolds on $so(8)$,'' {\em J. High Energy Phys.} {\bf 0212} (2002) 019, hep-th/0211003. 

\bibitem{Geom}
J.~de~Boer, M.~B. Halpern and C.~Helfgott, ``Twisted Einstein tensors and 
orbifold geometry,"
 {\em Int. J. Mod. Phys.} {\bf A18} (2003) 3489, hep-th/0212275.

\bibitem{Sch}
J.~Fr$\ddot{o}$hlich, O.~Grandjean, A.~Recknagel and V.~Schomerus, 
{\em Nucl. Phys.} {\bf B583} (2000) 381. 

\bibitem{Char}
V.~G. Kac and I.~T. Todorov, ``Affine orbifolds and rational conformal field theory extensions of $W_{1+\infty}$,''
    {\em Comm. Math. Phys.} {\bf 190} (1997) 57, hep-th/9612078; P.~Bantay, ``Characters and modular properties of permutation orbifolds,'' {\em 
  Phys. Lett.} {\bf B419} (1998) 175, hep-th/9708120; L.~Birke, J.~Fuchs, and C.~Schweigert, ``Symmetry breaking boundary conditions
  and {WZW} orbifolds,'' {\em Adv. Theor. Math. Phys.} {\bf 3}
   (1999) 671, hep-th/9905038; V.~G. Kac, R.~Longo, F.~Xu, ``Solitons in affine and permutation orbifolds,'' hep-th/0312512.

\bibitem{Del}
G.~W. Delius, ``Wess-Zumino-Witten model on discrete coset manifolds,'' 
{\em Phys. Lett.} {\bf B221} (1989) 283.

\bibitem{GKR}
M.~R. Gaberdiel, A.~O. Klemm and I.~Runkel, ``Matrix model 
eigenvalue integrals and twist fields in the su(2)-WZW model,'' hep-th/0509040.

\bibitem{Lep2}
B.~Doyen, J.~Lepowski and A.~Milas, ``Twisted vertex operators and 
Bernoulli polynomials,'' arXiv:math. QA/0311151 v2.

\bibitem{Orient1}
M.~B. Halpern and C.~Helfgott, ``Twisted open strings from closed strings: The WZW orientation orbifolds,"
 {\em Int. J. Mod. Phys.} {\bf A19} (2004) 2233, hep-th/0306014.

\bibitem{Orient2}
M.~B. Halpern and C.~Helfgott, ``On the target-space geometry of the 
open-string orientation-orbifold sectors," {\em Ann. of Phys.}
  {\bf 310} (2004) 302, hep-th/0309101. 

\bibitem{Basic}
M.~B. Halpern and C.~Helfgott, ``A basic class of twisted open WZW strings,"
{\em Int. J. Mod. Phys.} {\bf A19} (2004) 3481, hep-th/0402108.

\bibitem{Gentwopen}
M.~B. Halpern and C.~Helfgott, ``The general twisted open WZW string,'' 
{\em Int. J. Mod. Phys.} {\bf A20} (2005) 923, hep-th/0406003.

\bibitem{O's}
A.~Sagnotti, ``Open strings and their symmetry groups," {\em ROM2F-87/25}, talk presented at the Cargese Summer Institute
  on Non-Perturbative Methods in Field Theory, Cargese, Italy, July 
  16-30, 1987, hep-th/0208020; P.~Horava, ``Strings on world sheet orbifolds," {\em Nucl. Phys} 
  {\bf B327} (1989) 461; J.~Dai, R.~G. Leigh and J.~Polchinski, ``New connections among string theories,"
 {\em Mod. Phys. Lett.} {\bf A4} (1989) 2073; P.~Horava, ``Chern-Simons gauge theory on orbifolds: Open strings from 
three dimensions," {\em J. Geom. Phys} {\bf 21} (1996) 1, hep-th/9404101.

\bibitem{Giusto}
S.~Giusto and M.~B. Halpern, ``Hamiltonian formulation of open WZW strings," {\em Int. J. Mod. Phys.}
 {\bf A16} (2001) 3237, hep-th/0101220.
 
 \bibitem{DWZW}
A.~Yu. Alekseev and V.~Schomerus, ``D-branes in the WZW model," {\em Phys. Rev.} {\bf D60}, (1999) 061901,
 hep-th/9812193; J.~Fuchs and C.~Schweigert, ``Symmetry breaking boundaries I. General theory," 
  {\em Nucl. Phys.} {\bf B558} (1999) 419,  hep-th/9902132; G.~Felder, J.~Fr\"{o}hlich, J.~Fuchs and C.~Schweigert, ``The geometry of WZW branes,"
 {\em J. Geom. Phys.} {\bf 34} (2000) 162, hep-th/9909030; S.~Stanciu, ``D-branes in group manifolds,"
  {\em J. High Energy Phys.} {\bf 0001} (2000) 025, hep-th/9909163; C.~Bachas, M.~Douglas and C.~Schweigert, ``Flux stabilization of D-branes,"
 {\em J. High Energy Phys.} {\bf 0005} (2000) 048, hep-th/0003037; S.~Fredenhagen and V.~Schomerus, ``Branes on group manifolds, gluon condensates, and twisted K-theory," 
{\em J. High Energy Phys.} {\bf 0104} 007, hep-th/0012164.

\bibitem{Gawed}
K.~Gawedzki, I.~Todorov and P.~Tran-Ngoc-Bich, ``Canonical quantization of the boundary Wess-Zumino-Witten model,"
 hep-th/0101170.
 
\bibitem{TwWZW}
 A.~Yu Alekseev, S.~Fredenhagen, T.~Quella and V.~Schomerus, 
 ``Non-commutative gauge theory of twisted D-branes,'' {\em Nucl. 
 Phys.} {\bf B646} (2002) 127, hep-th/0205123; H.~Ishikawa and T.~Tani, ``Twisted boundary states and 
representation of generalized fusion algebra,'' hep-th/0510242; G.~Naculich and H.~J.Schnitzer, ``Level-rank duality of untwisted and 
twisted D-branes,'' hep-th/0601175.

\bibitem{Permbranes}
A.~Recknagel, ``Permutation branes,'' {\em JHEP} {\bf 04} (2003) 041,
hep-th/0208119; H.~Enger, A.~Recknagel and D.~Roggencamp, ``Permutation branes and
linear matrix factorizations,'' hep-th/0508053; I.~Brunner and M.~R. Gaberdiel, `` Matrix
 factorizations and permutation branes,'' {\em JHEP} {\bf 07} (2005) 012, 
 hep-th/0503207; S.~Fredenhagen and T.~Quella, ``Generalized permutation branes,'' 
hep-th/0509153.

\bibitem{Bars}
I.~Bars, C.~Deliduman and O.~Andreev, ``Gauged duality, conformal
symmetry and space-time with two times,'' {\em Phys. Rev.} {\bf D58} (1998) 
066004, hep-th/9803188.

\bibitem{Noghost}
P.~Goddard and C.~B. Thorn, ``Compatibility of the dual pomeron with 
unitarity and the absence of ghosts in the dual resonance model,'' {\em Phys. Lett.} 
 {\bf B40} (1972) 378.

\bibitem{DV2}
R.~Dijkgraaf, E.~Verlinde and H.~Verlinde, ``Matrix string theory,''
  {\em Nucl. Phys. } {\bf B 500} (1997) 43, hep-th/9703030.
 
\bibitem{Deser}
S.~Deser and B.~Zumino, ``A complete action for the spinning string,'' 
{\em Phys. Lett.} {\bf B65} (1976) 369.

\bibitem{R}
P.~Ramond, ``Dual theory for free fermions,'' {\em Phys. Rev.} {\bf D3} (1971) 2415.

\bibitem{NS}
A.~Neveu and J.~H. Schwarz, ``Factorizable dual model of pions,'' {\em Nucl. Phys.} {\bf B31} (1971) 86.

\bibitem{Schout}
P.~Bouwknegt and K.~Schoutens, ``W-symmetry in conformal field 
theory,'' {\em Phys. Rep.} {\bf 223} (1993) 183, hep-th/9210010.

\bibitem{Z}
A.B. Zamolodchikov, ``Infinite additional symmetries in 
two-dimensional conformal quantum field theory,'' {\em Theor. Math. 
Phys.} {\bf 65} (1985) 1205.

\bibitem{Yam}
M.~B. Halpern and J.~P. Yamron, `` A generic affine-Virasoro 
action,'' {\em Nucl. Phys.} {\bf B351} (1991) 333.

\bibitem{Lin'd}
J.~de Boer, K.~Clubok and M.~B.~Halpern, ``Linearized form of the generic affine-Virasoro action," 
{\em Int. J. Mod. Phys.} {\bf A9} (1994) 2451, hep-th/9312094.

\bibitem{Teit}
C.~Teitelboim, ``The Hamiltonian structure of two-dimensional 
space-time and its relation with the conformal anomaly,'' in Quantum 
Theory of Gravity, ed. S.M. Christenson (Adam Hilger, Bristol, 1984).

\bibitem{Polya}
A.~Polyakov, ``Quantum geometry of bosonic strings,'' {\em Phys. 
Lett.} {\bf B103} (1981) 207. 

\bibitem{Nambu}
Y.~Nambu, Lectures at the Copenhagen Summer Symposium, 1970, 
unpublished; T.~Goto, ``Relativistic quantum mechanics of one-dimensional mechanical
continuum and subsidiary condition of dual resonance model,''
 {\em Prog. Theor. Phys.} {\bf 46} (1971) 1560. 
 
\bibitem{GGRT}
P.~Goddard, J.~Goldstone, C.~Rebbi and C.~B. Thorn, ``Quantum dynamics
of a massless relativistic string,'' {\em Nucl. Phys.} {\bf B56} (1973) 109.

\end{thebibliography}
\end{document}